# Validation of Space Weather Forecasting at Leonid Space

*Rev 1.0, authored 2026-10-10 by Scott Shambaugh*
leonidspace.com/contact

## Executive Summary

Leonid Space's mission is to solve orbital drag, from days to decades. We generate real-time predictions of space weather indices, calibrated whole-Earth density grids, spacecraft orbits, and satellite lifetimes. This report validates our internal space weather forecasts through comprehensive cross-validation of retroforecasts scored against the entire F10.7 and Ap historical record. This is compared against all other publicly available forecast archives for these data, and shows Leonid's to be the most accurate available.

### Methodology

We built a unified forecast model that jointly forecasts F10.7 and Ap on all timescales from 3 hours up to multi-decade chains of future solar cycles. This was used to generate an archive of historical 15-year retroforecasts, after performing a leave-one-out fit of the model on the data outside those 15 years and conditioning only on historical data available at the time. The accuracy of this archive was scored against measured space weather indices at all lookahead horizons from 1 day to 15 years. All other publicly available space weather forecast archives were also scored using the same approach.

### Key Results

**Leonid shows the best F10.7 and Ap prediction accuracy of any space weather forecast model archive, across all lookahead time horizons.** This includes 15 F10.7 forecasts and 12 Ap forecasts from NASA's MSAFE, SWPC's 27-day, 45-day, and Solar Cycle predictions, SWPC/USAF's RSGA, SIDC's Ursigram, CLS, Athens, KSO, BGS, SEPC's 3-day and 27-day forecasts, AFRL/NSO's SIFT, BoM's 3-day and Cycle predictions, and NASA's Schatten forecasts. For Ap it is strictly better across all time horizons. For F10.7 it gives strictly better accuracy up to 5 years into the future, and is tied with MSAFE beyond that. Performance is best realized for workflows that can use Leonid's calibrated forecast uncertainty.

### Implications

These results establish a performance baseline for Leonid's space weather predictions. The demonstrated accuracy supports collision avoidance, target tracking, mission planning, and operational decision-making for LEO & VLEO satellite operators facing an increasingly dynamic space environment.

## Document History

| Revision / DOI | Prepared By | Software version | Changes | Published |
|---|---|---|---|---|
| Rev. 1.0 (this document) | Scott Shambaugh | 2.6.0 | - | 2026-10-10 |
| | | | | |
| | | | | |

## Overview

Leonid Space exists to solve orbital drag, from days to decades. We focus on the VLEO and LEO space environments, where drag dominates satellite motion. In the "tactical" time horizon of hours-to-days, this information is critical for target tracking and collision avoidance. In the "strategic" time horizon of years-to-decades this drives satellite lifetime predictions, which are needed for fleet replenishment planning, revenue forecasts, and mission continuity. We want operators to know while they can still act.

The goal of Leonid's space weather forecasting is to generate an ensemble of realistic F10.7 and Ap predictions that accurately bound future behavior under calibrated uncertainty. This in turn drives our atmosphere model and satellite drag predictions. Our atmosphere model is based on NRLMSISE-00, which takes in F10.7 and Ap as its space weather inputs. Other physics-based atmosphere models could take in additional driving parameters, however we prefer to use a fast, simple base model that we then correct using real-time data assimilation.

Moving from public space weather forecasts to an internal predictive model was originally motivated by improving our long-duration deorbit predictions. Our predictive pipeline validation analysis [1] included a variance decomposition step which showed that ~80% of the variance in our lifetime prediction error was attributable to the MSAFE solar flux forecasts that we were using. In building decadal-scale forecasts to replace MSAFE, we found that the techniques we were using could be extended to the near term as well, and saw strong performance improvements once we switched our near-term orbit predictions to a unified internal model.

This report shares the results from scoring our forecasts, and presents a direct apples-to-apples comparison against all other publicly available space weather forecast archives. This is also the most comprehensive cross-model comparison we are aware of. We show through a cross-validation approach that Leonid's F10.7 and Ap forecasts both improve upon or match all other forecast models. Ours has the best prediction error over all time horizons, the fastest Ap update cadence, and the longest validation record.

## Space Weather Historical Data

The NRLMSISE-00 atmosphere model uses daily F10.7 solar flux and the 3-hourly geomagnetic Ap indices as the driving space weather inputs to its density calculations. These are both obtained from the datafiles published by Geomagnetic Observatory Niemegk, GFZ Helmholtz Centre for Geosciences [2]. Flux data is available since 1947-02-14, and Ap data is available since 1932-01-01. All forecasts were converted to and scored against F10.7 observed at Earth, rather than adjusted to 1 AU.

Prior to this, sunspot number (SSN) counts are available from SILSO (Sunspot Index and Long-term Solar Observations) at the Royal Observatory of Belgium. The record of daily counts goes back to 1818, and monthly counts back to 1749. From the overlap in SSN and flux data from 1947 to now, we see that the relationship between SSN and F10.7 is strongly correlated, with [3] fitting a popular $4^{th}$ order polynomial to the relationship. We replicate the report by [3] of an apparent step change in flux measurements in 1981, however without a confirmed mechanism for the change we believe that there are too few cycles to rigorously assert this transition and fall back to the principle of parsimony in applying a single fit over all eras. Instead of a polynomial we prefer a "hockey stick" form inspired by [4], which scores the same with 3 parameters instead of 5, settles at the quiescent floor, and does not diverge at high SSN. Here $F_0$ is the flux floor, $S_0$ is the shoulder SSN, and $a$ is a linear scale between flux and SSN. We additionally model the probabilistic spread of this relationship.

$$F_{10.7} = F_0 + a * SSN\left(1 - e^{-SSN/S_0}\right)$$

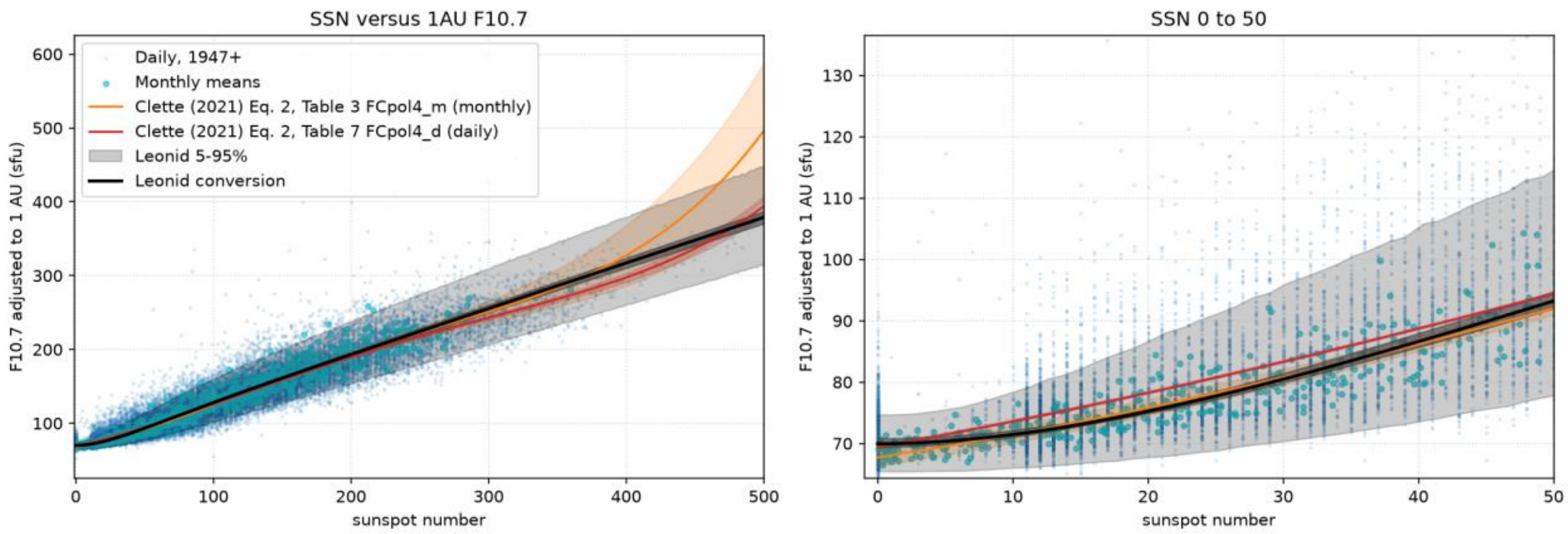

Figure 1: Clette 4th order polynomial fits and Leonid's hockey stick fit. Shaded bands show the 90% confidence interval for the central fit, and Leonid's 90% prediction interval is additionally shaded.

For geomagnetic data prior to the Ap cutoff in 1932, the National Centers for Environmental Information archives the geomagnetic aa index from 1868-01-01 to 2010-12-31. There is strong correlation here as well, however data discretization from rounding (notably causing aa to have a minimum of 2 instead of 1) and the nonlinearity of the relationship mean that we do not have a clean analytical fit. Here we used an empirical piecewise linear fit to estimate the Ap index from 1868-1932 aa data.

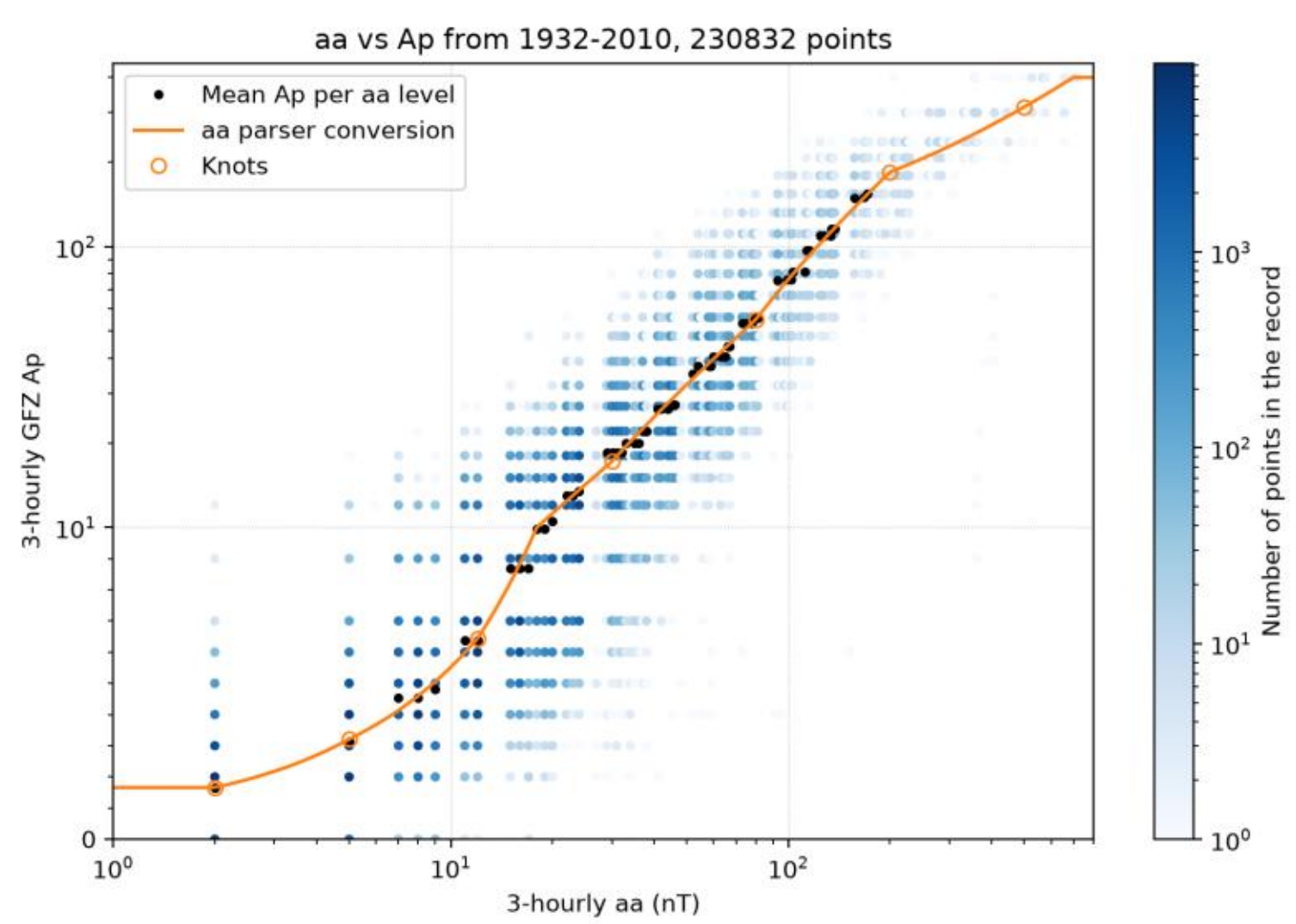

Figure 2: aa vs Ap from 1932 to 2010, with Leonid's empirical piecewise linear fit superimposed.

## Solar Cycle Behavior

After these two proxy variable conversions, our daily F10.7 record goes back to 1818 at the tail end of Solar Cycle 6 (18.5 full cycles), and our Ap record goes back to 1868 at the start of Cycle 11 (14 full cycles). Figure 3 shows how cycles compare, with starts measured from their 13-month smoothed SSN minima.

Several characteristics are worth noting. Flux tends to peak quickly and decay slowly. Geomagnetic activity is somewhat offset from the flux peak, with its maximum strength expected during the declining phase of the cycle. The peak strength of each cycle for both F10.7 and Ap is highly variable, but remains persistently high or low within each cycle. And the duration of each cycle is also variable, well approximated by a normal distribution of 11.03 ± 1.16 years.

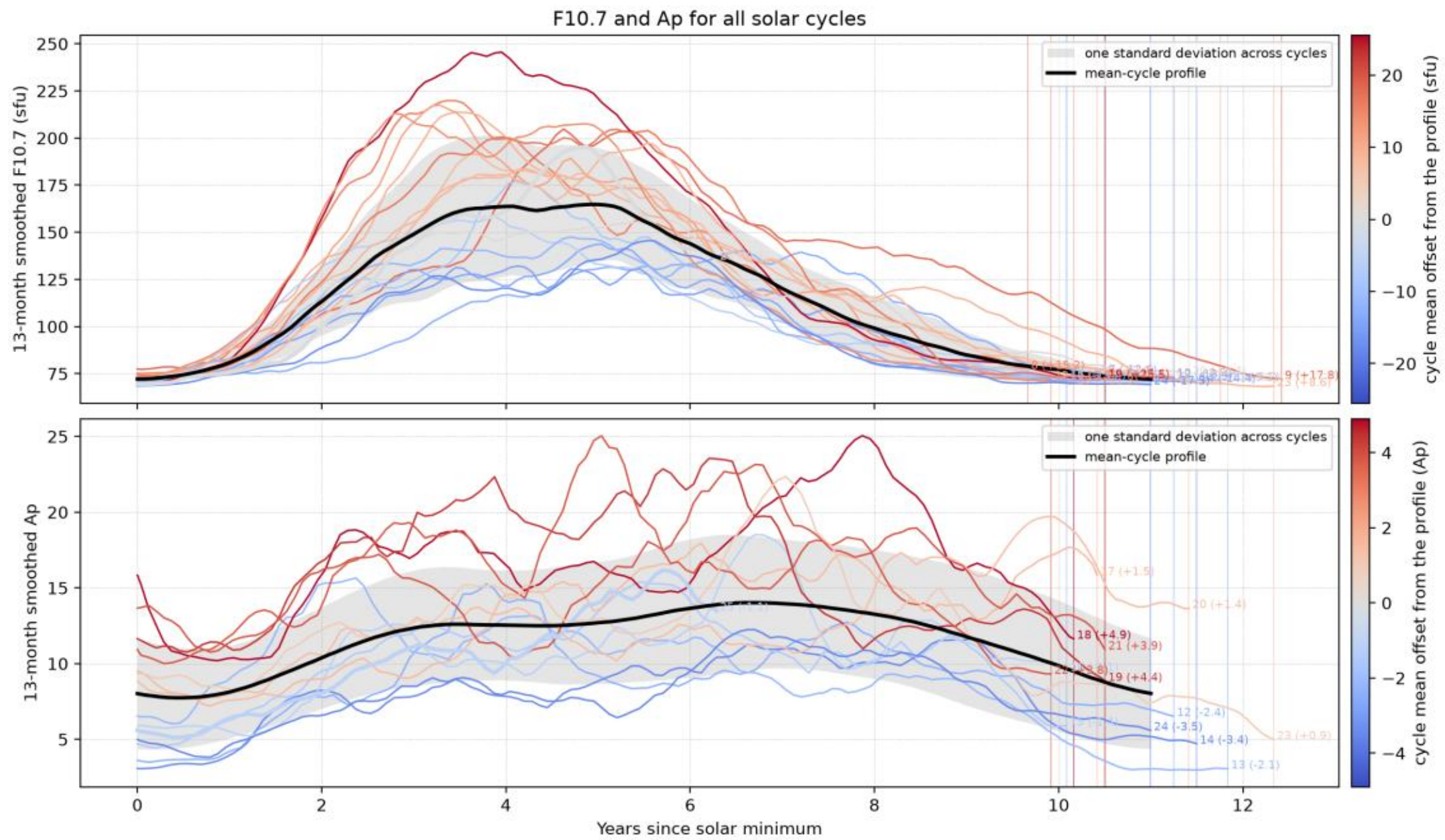

Figure 3: Smoothed F10.7 flux (top) and geomagnetic Ap (bottom) overlaid for all historical solar cycles. Black shows the average 13-mo smoothed cycle, and others are colored by their average offset from that mean. Shaded area shows one standard deviation spread.

## Public Space Weather Forecasts

We collected all F10.7 and Ap forecasts with a history of publicly available predictions. In addition to the sources listed below there are a number of individual or discontinued forecasts of solar cycle behavior. Notable examples include many academic papers, the 2019 NASA/NOAA Solar Cycle 25 Prediction Panel consensus [5], and SWPC's discontinued cycle predictions from its weekly PRF reports. We also omit Kp-only forecasts which are convertible to Ap, focusing only on products that forecast Ap directly [2].

We consider scoring one-off forecasts to be meaningless. Any successful $n = 1$ prediction cannot be disentangled from luck. It is instead necessary to score collections of predictions in order to validate the predictive power of the *forecast model*. Large collections of contemporary predictions are unfortunately not common. There are a number of forecast models which we would like to have scored, but did not have publicly available archives. This includes government and academic forecasts from IZMIRAN in Russia [6], Institute of Ionosphere in Kazakhstan [7] [8], GFZ's 3-day Ap [9] [10], BGS's cycle-scale F10.7 [11], KSO's 1-24 month F10.7 [12], ESA's SOLMAG [13] [14], and NASA's Debris Assessment Software forecasts [15]. It also includes commercial forecasts such as Space Environment Technologies' F10.7 predictions [16] [17]. We use what is publicly available, and urge space weather predictors to issue retroforecast archives that can be used to rigorously evaluate their generating models. Figure 4 and Figure 5 show the available archive coverage for each forecast model.

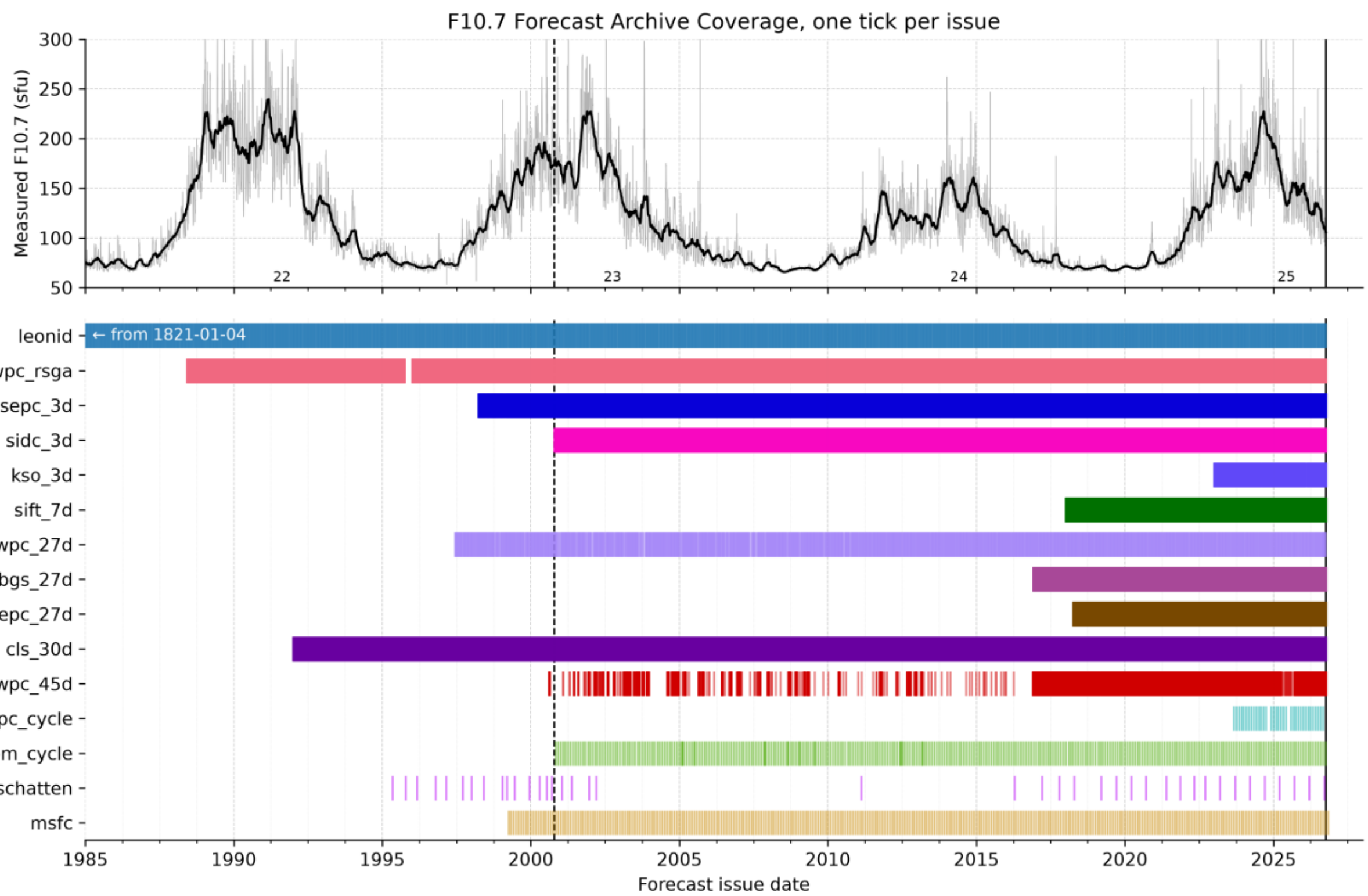

Figure 4: Historical F10.7 forecast archive coverage for each space weather forecast model (bottom), and corresponding measured flux (top). The vertical line marks our analysis start date.

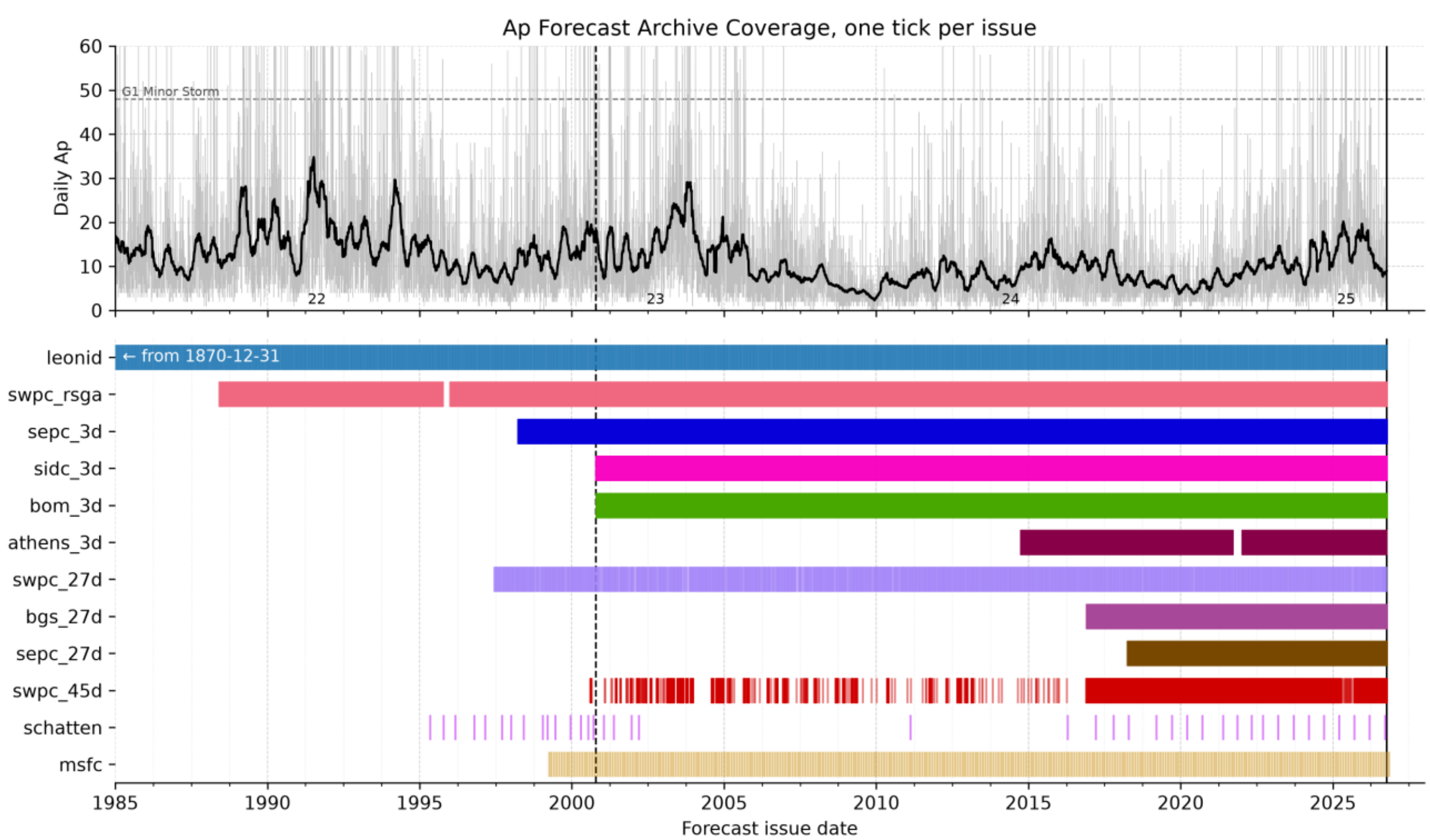

Figure 5: Historical Ap forecast archive coverage for each space weather forecast model (bottom), and corresponding measured Ap (top). The vertical line marks our analysis start date.

Taken together, these 15 flux forecasts and 12 Ap forecasts compared on common lookahead metrics make up the most comprehensive cross-comparison of space weather forecasts we are aware of, with [18]'s comparison of 7 Ap models being the closest we can find.

### SWPC/USAF Report of Solar-Geophysical Activity (RSGA)

The RSGA is a daily space weather forecast put out jointly by NOAA's Space Weather Prediction Center (SWPC) and the US Air Force (USAF). It provides 3-day forecasts of F10.7 and Ap, along with descriptions of recent solar activity and probabilities for storm events [19]. Text records extend back to 1988-06-02 [20], and scanned pdfs extend back to 1966. We performed OCR on the scans, but do not include them in our test set due to some open questions on measurement equivalence between the eras.

### SWPC 27-Day Forecasts

SWPC (formerly NOAA's Space Environment Center / SEC) has released 27-day forecasts of F10.7 and Ap in its weekly "Preliminary Report and Forecast of Solar Geophysical Data" (PRF) report since 1975-09-09 [21]. However, digital records only go back to 1997-06-10 [22]. Since roughly 2012, ~90% of the 1-3 day forecasts have been identical to the RSGA, suggesting a shared model.

### SWPC/USAF 45-Day Forecasts

SWPC recently began releasing the 45-day forecasts that were previously being generated for the USAF. Unfortunately, earlier forecasts do not have a complete public archive, and so we combined what is available since the handover on 2026-03-02 [23], with 1119 issues that were recoverable from Internet Archive mirrors of SEC/SWPC's FTP server [24], and 2843 issues from ESA's FORIND archive [25]. The 1-7 day forecast horizons for F10.7 and Ap are regenerated daily, and the 8-45 day outlooks are updated weekly [26].

The 45-day forecast is in large part a composite of other forecasts. The 1-3 day F10.7 and Ap values match the RSGA for that day in ~95% of issues, and the 8-27 day window matches the SWPC 27-day forecast in ~90% of issues, with the rest of the data generated some other way. In mid-2025 the USAF product switched from manual forecasting files to an automated "SET4D system", which for the 28-45 day window simply repeated the 1-27 day data. The switch to SWPC ownership in 2026 no longer shows this automated repeat of data in the 28-45 day window.

### SIDC's Ursigram

The Solar Influences Data Analysis Center (SIDC) at the Royal Observatory of Belgium has posted a daily "Ursigram" bulletin since 2000-10-16 with 3-day F10.7 and Ap forecasts [27] [28].

### CLS's Solar Flux Forecast

Collecte Localisation Satellites (CLS) and CNES have collaborated to release daily 30-day F10.7 forecasts since 1992-01-01 [29]. Note that data prior to May 2024 appears to all have been generated at dates after their forecast start, which we think represents a retroforecast. See [30] for methodology.

### AFRL/NSO's SIFT

The Air Force Research Laboratory (AFRL) and National Solar Observatory (NSO) co-developed the Solar Indices Forecasting Tool (SIFT), which uses magnetogram observations from the Global Oscillation Network Group (GONG) to generate 7-day forecasts of F10.7 [31] [32] [33].

### Athens' 3-day Forecasts

The Athens Space Weather Forecasting Center, run by the Cosmic Ray Group at the National and Kapodistrian University of Athens, has released a daily forecast of 3-day Ap since 2014-10-01 [34] [18].

### KSO's 3-day Forecasts

Kanzelhöhe Observatory (KSO) at the University of Graz, together with Skoltech, has released 1-3 day F10.7 forecasts since 2022-12-30, distributed through ESA's Space Weather Service Network [35] [36] [37]. The ESA product page lists F10.7 uncertainty as 5.78 sfu for 1 day ahead, 8.79 sfu for 2 days ahead and 11.62 sfu for 3 days ahead, which we assume represents one standard deviation.

### SEPC's 3-day Forecasts

The Space Environment Prediction Center (SEPC / 空间环境预报中心) at the Chinese Academy of Sciences has released a daily forecast of 3-day F10.7 and Ap since 1998-03-25 [38].

### SEPC's 27-day Forecasts

SEPC has also released a 27-day forecast of F10.7 and Ap since 2018-04-02 [39] [40]. Note that the first 3 days are identical to the 3-day forecast in >99% of cases.

### BGS's 27-day Forecasts

The British Geological Survey (BGS) has released a daily 27-day forecast of F10.7 and Ap since 2016-11-26 [41], archived by ESA's Space Weather Service Network [25].

### BoM's 3-day Forecasts

The Australian Space Weather Forecasting Centre at the Bureau of Meteorology (BoM) produces a daily Solar and Geophysical Report, which forecasts the next 3 days of F10.7 and Ap [42]. However, we see that its F10.7 forecasts shared >99% of values with the RSGA, and we skip scoring those under the assumption that its flux predictions are simply a redistribution of RSGA's. The Ap forecasts appear to be independent, so we score those.

### BoM's Cycle Forecasts

BoM also releases a monthly Observed and Predicted Solar Indices bulletin, which forecasts the predicted monthly F10.7 for the rest of the cycle [43]. This has archives dating back to 2000-11, but is based on 13-month smoothing. Prior to 2007-11, 12 months of these forecasts came with an associated uncertainty, however this information is not included in later forecasts. We assume this uncertainty represents one standard deviation.

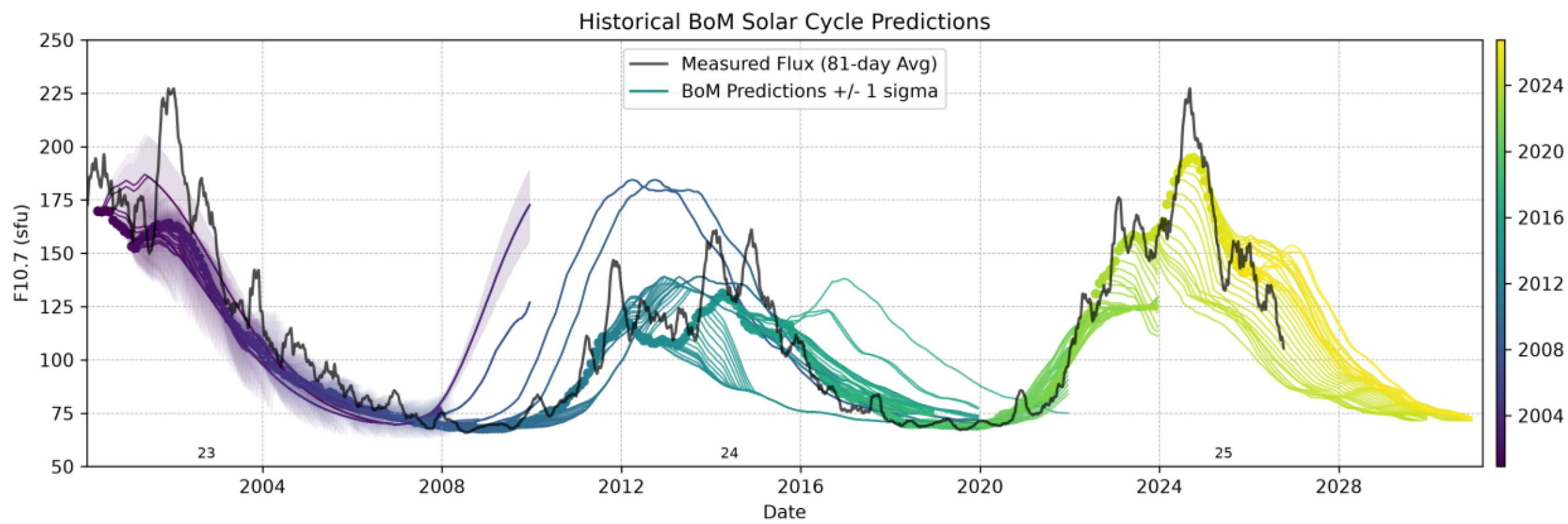

Figure 6: All historical solar cycle predictions from BoM's bulletin, colored by prediction date and with their ±1 standard deviation uncertainty range shaded. Uncertainty intervals stopped being published in 2007-11. The adjusted 81-day average F10.7 solar flux is overlaid in black.

## SWPC Solar Cycle Predictions

SWPC overhauled its solar cycle prediction model in 2023, and has released monthly forecasts through the upcoming solar cycle since 2023-09 [44] [45] [46] [47]. Like the MSAFE model, this is built on 13-month smoothing of flux and loses information at shorter timescales. Note that the current data on SWPC's website only extends through the end of Solar Cycle 25 in 2030-12 [48].

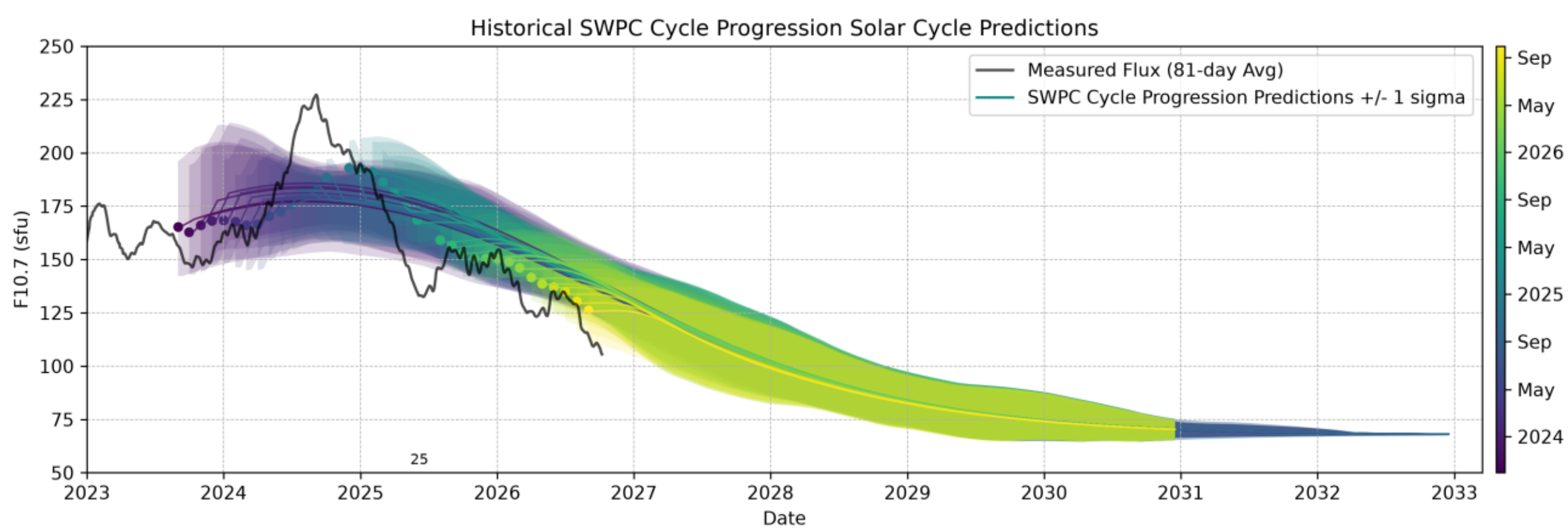

Figure 7: All historical solar cycle predictions from SWPC's Cycle forecast, colored by prediction date and with their ±1 standard deviation uncertainty range shaded. The adjusted 81-day average F10.7 solar flux is overlaid in black.

## NASA MSFC's Solar Activity Future Estimation (MSAFE)

The MSAFE forecasts published by NASA's Marshall Space Flight Center (MSFC) have continuous monthly forecasts archived since April 1999 [49] [50] [51]. It uses a modified McNish-Lincoln method to predict one solar cycle into the future (currently through 2041) [52] [53]. We disperse these by calculating a split-normal distribution from the 5$^{th}$ and 95$^{th}$ percentile forecasts in the datafiles [49]. The MSAFE forecast model is based on 13-month smoothing of solar flux, which causes it to lose all information at shorter timescales. This limits usefulness for the days-to-weeks timescale.

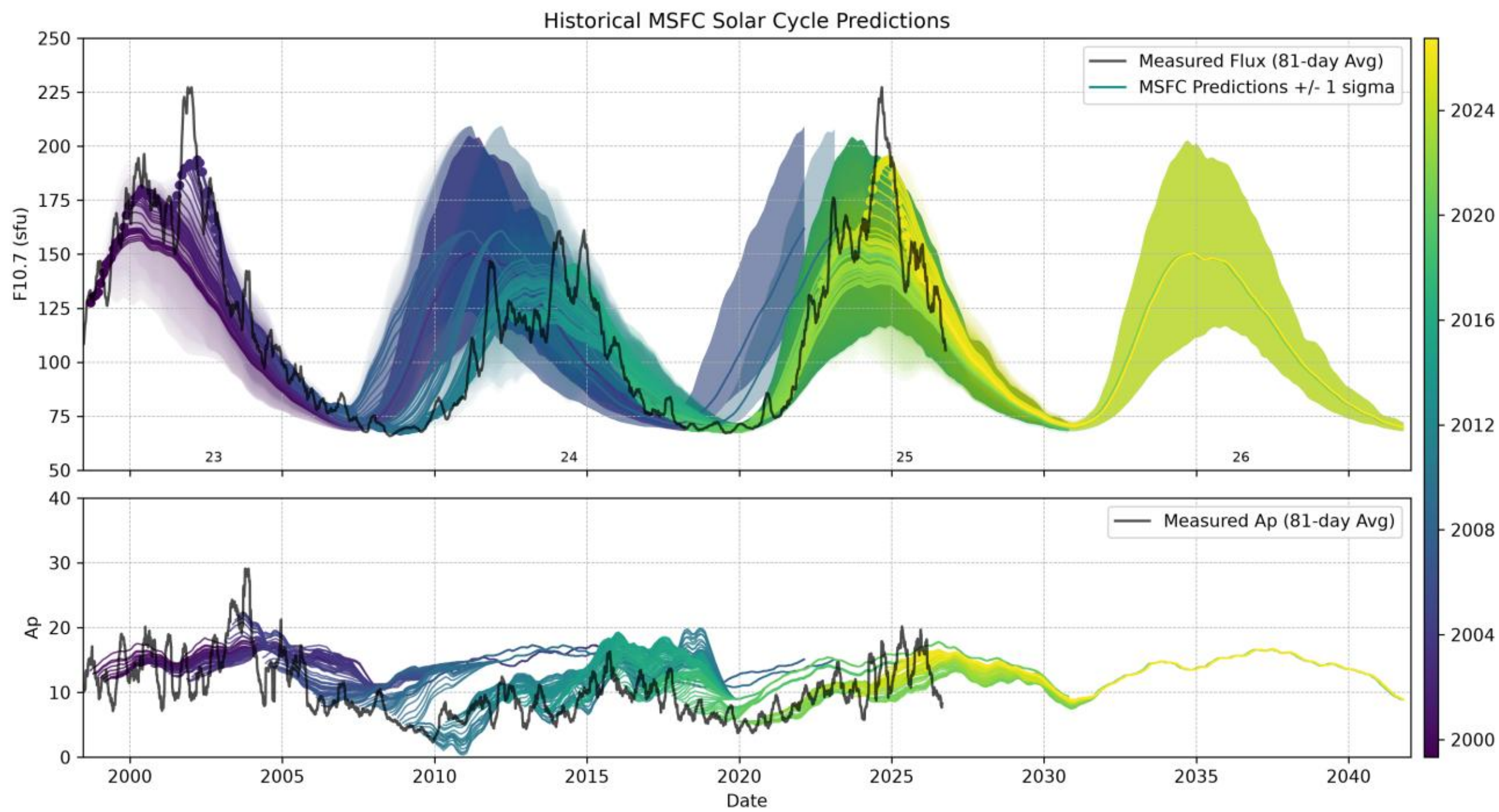

Figure 8: All historical F10.7 (top) and Ap (bottom) solar cycle predictions from NASA's MSAFE, colored by prediction date and with their ±1 standard deviation uncertainty range shaded. The adjusted 81-day average is overlaid in black.

## NASA GSFC's Schatten Forecasts

The NASA GSFC Flight Dynamics Facility issues the Schatten solar cycle predictions every 6 months, dating back to at least 1995-05 [54] [55] [56]. NASA's iSWA archive only contains records since 2019-03, but 25 additional scattered issues were able to be found in the Internet Archive [24].

These provide nominal and ±2 sigma values for monthly F10.7 averages, for early, nominal, and late solar cycle timing. We could not find guidance on how to use the split timing for operational forecasts. In this analysis we took an ensemble approach that randomly chose nominal/early/late cycle data with 0.75/0.125/0.125 odds, and drew the F10.7 from a split-normal distribution defined by the uncertainty range for that case. Note that we found the Schatten scores to be incredibly noisy due to their low number of only 20 independent windows, which was obscuring the rest of the data on the scoring plots. So they are included in the scoring tables, but excluded from the plots for clarity.

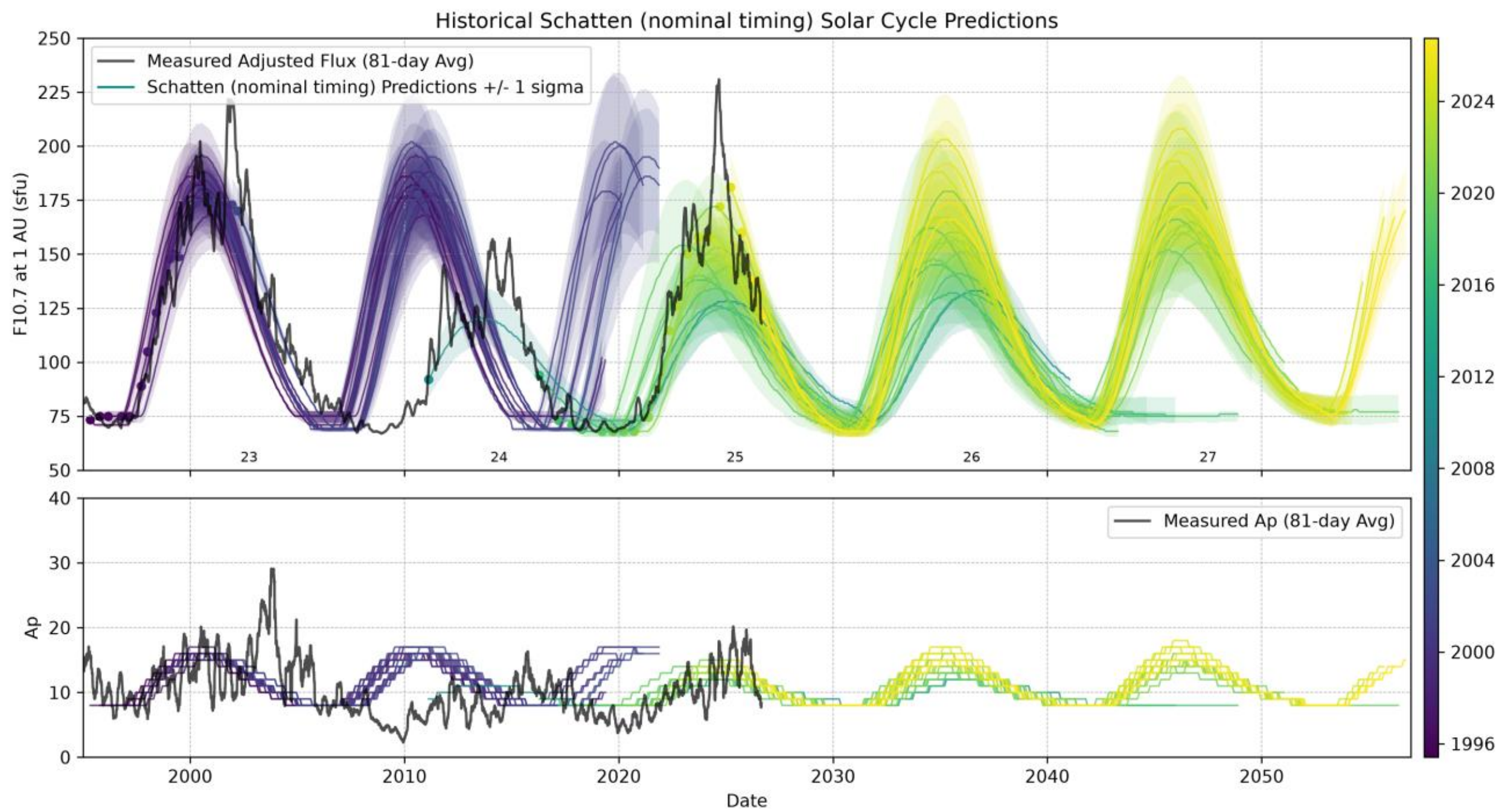

Figure 9: All historical F10.7 (top) and Ap (bottom) solar cycle predictions from NASA GSFC's Schatten forecasts, colored by prediction date and with their ±1 standard deviation uncertainty range shaded. The adjusted 81-day average F10.7 solar flux is overlaid in black. Only nominal timing predictions are shown, and this is an incomplete record with a large data gap from 2003-2016.

## Leonid Space Forecasts

Leonid Space has created internal predictive space weather forecasts for F10.7 flux and geomagnetic Ap, relying only on the historical record. Our unified generative model reconditions on measured data every 3 hours, jointly conditions on and forecasts F10.7 and Ap, and accounts for behavior at all timescales from 3 hours to multiple decades through chained cycles. It accounts for variations in cycle strength, shape, and duration, which is particularly important for decadal-scale missions where the possibility of chained worst-on-worst or best-on-best cycles are not captured by any other forecast model. Leonid's forecasts are probabilistic, and intended to be used via Monte Carlo sampling of future scenarios, as shown in Figure 10. We do not currently use storm-specific data streams to try to forecast solar flares, CMEs, or other events which cause geomagnetic storms, though these events are still forecasted probabilistically. Assimilation of additional data streams will be the focus of a later effort.

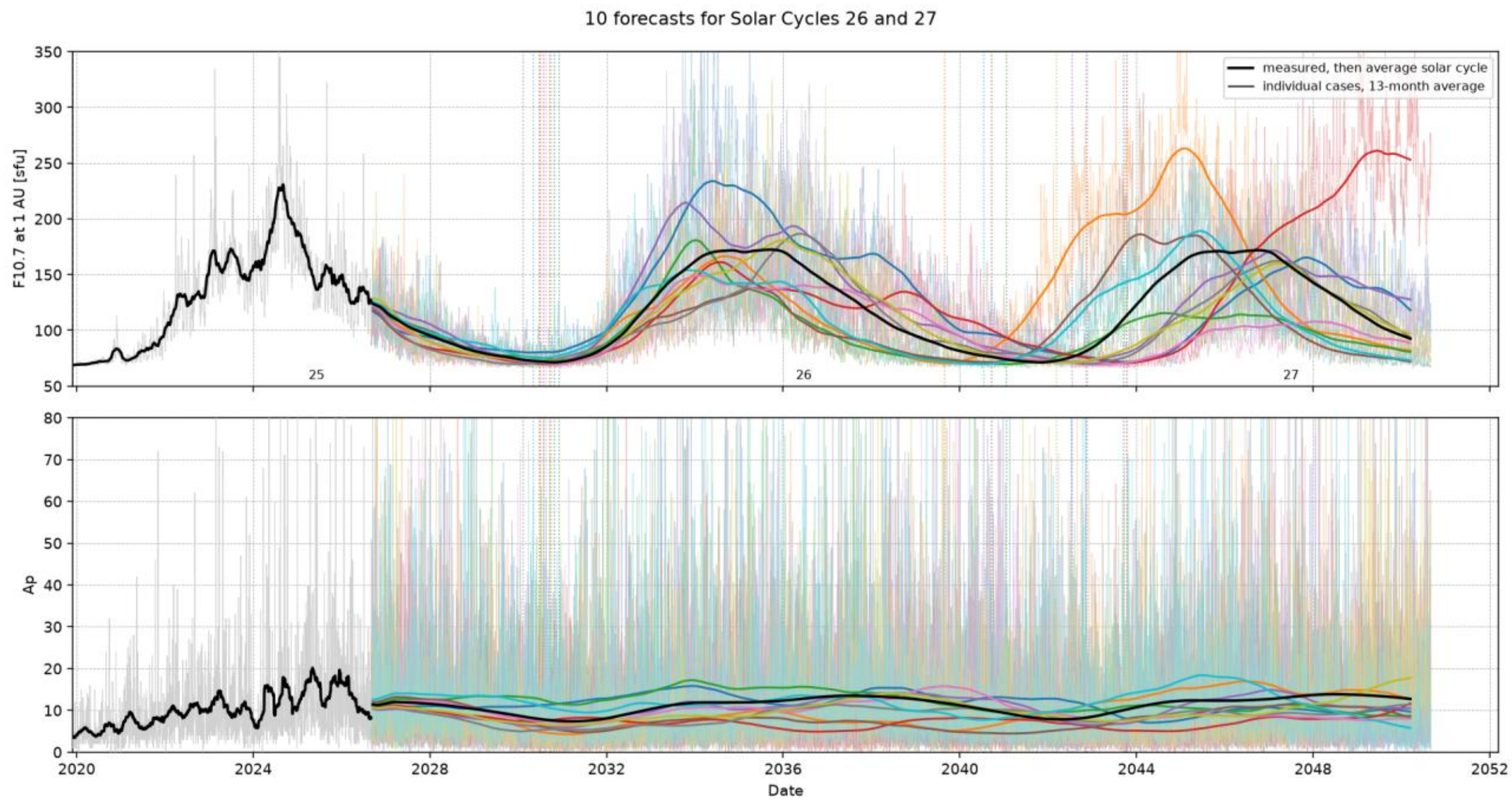

Figure 10: Ten randomly sampled forecasts for Solar Cycles 26 and 27 from Leonid's space weather forecast model, showing F10.7 (top) and Ap (bottom). The average cycle is overlaid in black.

The architecture and predictive approach of our forecast model is proprietary and will not be shared here. However, we can share its outputs and the results of our cross-validation effort. Figure 11 below shows our nominal predictions for all historical cycles, generated by dropping that cycle from the historical dataset and fitting our model on the rest of the record. This heuristically shows how our model adapts to measured conditions as the solar cycle progresses, and we score this model rigorously against other forecast models in the next section.

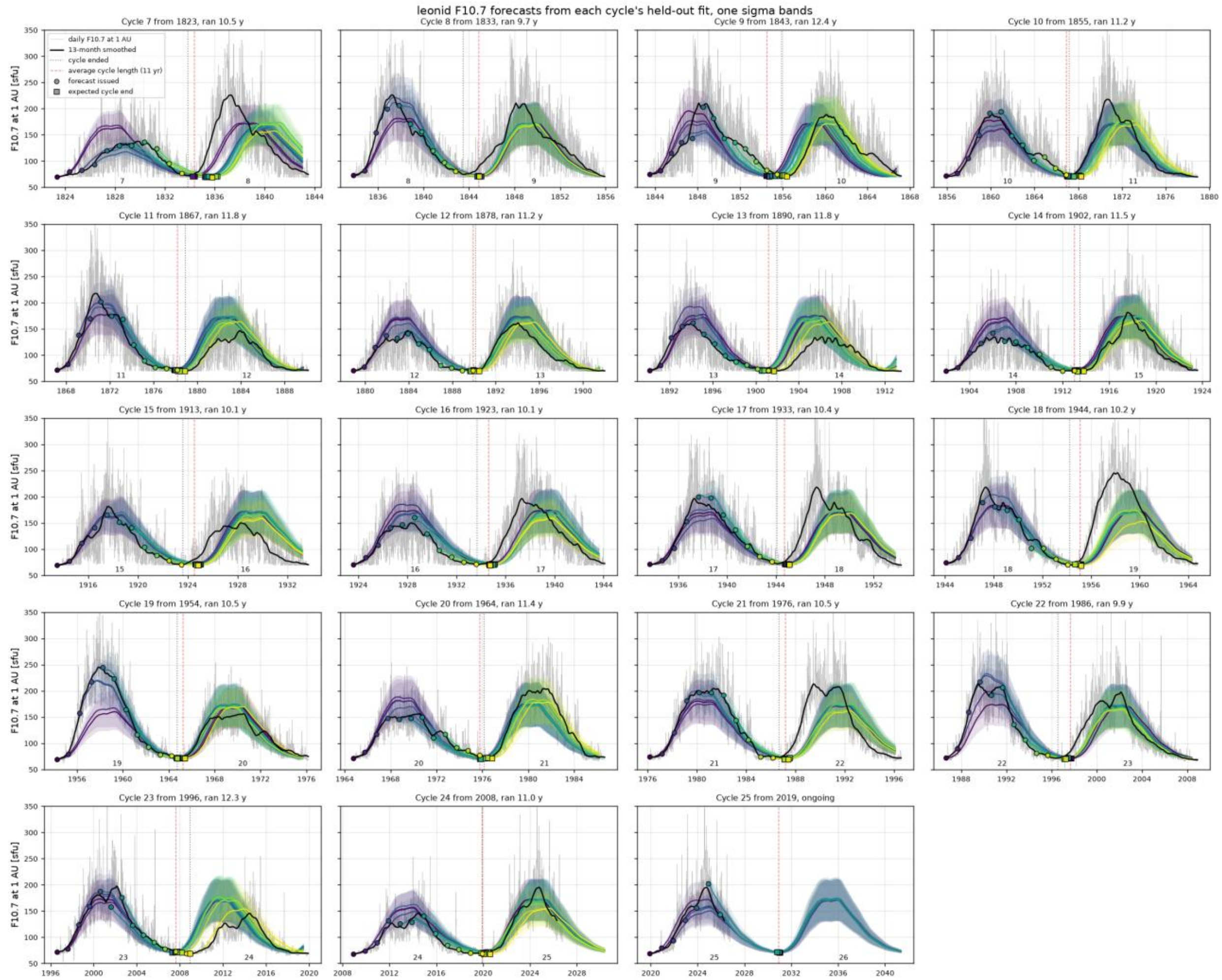

Figure 11: Leonid's rolling yearly forecasts for Solar Cycles 7-25, from a leave-one-out model fit that excludes the solar cycle in question. These show nominal forecasts and uncertainty, rather than a range built from Monte-Carlo draws. Circles indicate the forecast date, squares the estimated cycle end date, and shaded regions the 1-sigma prediction interval. Colors indicate the time within each cycle that the forecast was issued. One cycle ahead is included to show the effect of conditioning on nominal timing and amplitude – note that uncertainty in timing remains wide until the rising phase of the next cycle begins.

## Scoring

We score collections of forecasts against each lookahead duration that they predict. For example, all 1-day predictions for a given forecast are aggregated and compared against the realized historical values to measure the forecast error at 1 day. Forecast horizons up to 45 days are scored at a 1-day interval, up to 180 days at a 7-day interval, and up to 15 years at a 30-day interval. 15 years is chosen as a long-duration interval that covers a full cycle through to approximately the next solar max.

### Scores

The predicted minus true value for each forecast form the set of prediction errors for each time horizon, where positive error means that the value was lower than predicted. The mean bias (mean signed error) of all samples is calculated for each forecast as a measure of the prediction bias. The prediction accuracy is measured using an empirical continuous ranked probability score (CRPS) from the set of errors, which is the primary quality metric. The CRPS can be thought of as a generalization of the mean absolute error (MAE) for probabilistic forecasts, and for a single non-dispersed prediction is equivalent to MAE.

If we have the set of predicted values $x \in \{x_1, \dots, x_n\}$ and the true value $y$, then the CRPS value for each forecast horizon is calculated as below. Over the whole ensemble of $m$ forecasts we look at the mean of the CRPS values and mean errors to score the run. For this analysis we use $n = 200$ ensemble forecasts. The accuracy of Leonid's forecasts can also be measured using a CRPS Skill Score (SS), where a score above 0 and closer to 1 shows our predictive power relative to other sources. The skill score for source $s$ is calculated at each lookahead over the number of blocks $B_s$ at that lookahead, with each block holding $n_b$ individual forecasts.

$$CRPS = \frac{1}{n}\sum_{i=1}^{n}|x_i - y| - \frac{1}{2n(n-1)}\sum_{i=1}^{n}\sum_{j=1}^{n}|x_i - x_j|$$

$$SS = 1 - \frac{\bar{C}_{leonid}}{\bar{C}_{ref}} \qquad \overline{C_s} = \frac{1}{B_s}\sum_{b=1}^{B_s}\left(\frac{1}{n_b}\sum_{i\in b} CRPS_{s,i}\right)$$

Additionally, we score the dispersed predictions by calculating the rank percentile of each true value within the set of predictions for each forecast at each horizon. We then look at the ensemble of rank percentiles for all forecasts at this horizon and compare this distribution to an ideal uniform distribution, which we can plot as a histogram or a calibration curve of its empirical cumulative distribution function (ECDF). This is scored using the Root Mean Square Calibration Error (RMSCE), which measures the root mean squared difference between the ECDF and the straight-line CDF of the uniform distribution. The RMSCE can be interpreted as the average percentile error of the output distribution. The equations below show the formulas, where $p_k$ is the rank percentile of the true value for the $k^{th}$ forecast and $p_{(1)} \le \cdots \le p_{(m)}$ are those values when sorted. Percentiles below 0.5 indicate that the forecast predicted a higher value than reality. A perfect point forecast can only produce an even split between percentiles 0 and 1, which puts a lower bound of $RMSCE = \sqrt{1/12} \approx 0.2887$ on any forecast without associated uncertainty bands.

$$p_k = count\{x_{k,i} \le y_k\}/n$$

$$RMSCE = \sqrt{\frac{1}{12m^2} + \frac{1}{m}\sum_{k=1}^{m}\left(p_{(k)} - \frac{2k-1}{2m}\right)^2}$$

We have low confidence in lookaheads with fewer than 10 independent windows, and data there should be taken to be provisional. We drop scoring data altogether when it falls to 2 independent windows, since it is impossible to then generate an uncertainty band.

### Uncertainty Bands

At each lookahead for each forecast, we can calculate the uncertainty of the central estimate through a block t-interval approach [57]. Consecutively issued forecasts have overlapping forecasts, so by splitting the record into windows that are each one lookahead long we can treat these blocks as independent. The bands plotted in Figure 12 and Figure 15 are the 68% Student-t interval of the per-block means, which allows for appropriate widening of uncertainty as the number of blocks decreases. Uncertainty is undetermined when the lookahead is long enough such that only two blocks exist in the record.

Skill score uncertainties run into the instabilities inherent when calculating the uncertainty in a ratio of values, and are resolved using Fieller's confidence intervals under an independence assumption [58].

### Leave-one-out Cross-Validation

Leonid's historical archive was generated through retroforecasting, which requires us to screen out information which was not available at the time to avoid information leakage. In order to generate a robust model fit for scoring early cycles, we use a leave-one-out (LOO) cross-validation approach over a long 16-year window, rather than a leave-future-out (LFO) approach. For example, a Leonid forecast date in 1980 drops 16 years of data from 1980-1996, fits the model on the remaining data, and scores its 15-year forecast horizon against the historical data from 1980-1996. It still conditions on the full historical record prior to its forecast date. For forecast horizons under 1 year, we drop 2 years of data rather than 16.

The scores we report for Leonid at each time horizon are the average score over every forecast, as generated using its own LOO model fit. This ensures statistical robustness and no information leakage. We assume that all other forecast models are either real-time forecasts, or are similarly retroforecasts that have properly screened out their published intervals from their model fits.

### Lookahead Time Horizon

Scores are taken from all predictions since the start of SIDC's record on 2000-10-16 so that there is a consistent baseline set across most forecast models, while still providing a large 26-year history of solar behavior. Without this commonality, differences in scoring would instead be driven by the characteristics of the particular solar cycles under investigation in the long term. The SWPC Cycle, Schatten, SIFT, SEPC 27-day, KSO, BGS, and Athens forecast models have limited archive coverage, and may not be directly comparable.

### Daily vs Monthly Forecasts

For forecasts that predict monthly values, we pin that value to the $15^{th}$ of each month and linearly interpolate onto a daily grid, which we then score.

## F10.7 Forecast Comparisons

Table 1 shows a summary of each of the F10.7 forecast models, along with their MAE & CRPS at selected near-term time horizons. Leonid's forecast has the lowest CRPS predictive error at all time horizons, followed closely by CLS's and SIFT's forecasts over the short 1-3 day lookaheads. Note also that Leonid's F10.7 forecast is coupled with its Ap forecast, which neither SIFT nor CLS generates. Figure 13 shows the 1-day forecast error over time, demonstrating that the absolute magnitude of this predictive accuracy scales as a function of the solar cycle, but that Leonid's relative performance remains leading during both solar minimum and maximum. Remarkably, only a few forecast models outperform the persistence baseline for a 1-day F10.7 lookahead: Leonid, SEPC, SIDC, SIFT, and CLS.

Figure 12 shows all scores for all models as a function of lookahead. Leonid's excellent CRPS performance extends smoothly over all time horizons, matching or exceeding the performance of all other forecast models at all lookaheads. We show parity with MSAFE beyond the ~5-year time horizon. Part of this excellent CRPS scoring comes from its well-calibrated uncertainty over all time horizons, as shown by its low RMSCE. However even the MAE of its point forecasts is best-in-class. The bias plot shows most models to be unbiased up to the 1-year lookahead. The drop to single-digit number of independent windows past the ~2.5 year time horizon means that behavior past this time horizon is largely characteristic of the behavior of solar cycles 23, 24, and 25, and does not generalize well.

Table 1: Comparison of F10.7 forecast models, along with a persistence baseline. The CRPS of a point forecast is equal to its MAE, so those scores are shown as one where no uncertainty quantification exists. Bolding shows the best in each column, and asterisks indicate scores over only a portion of the record.

| F10.7 Forecast Model | Archive Start | Update Cadence | Forecast Horizon | Resolution | Quantified Uncertainty | Lookahead F10.7 MAE / CRPS, 2000-10-16 onwards (SFU) | | | | |
|---|---|---|---|---|---|---|---|---|---|---|
| | | | | | | 1-Day | 3-Day | 7-Day | 27-Day | 1-Year |
| Leonid | **1821-01-04** | 3-Hourly | **Unbounded Multi-Cycle** | **Daily** | **Yes** | 3.8 / **2.9** | 7.2 / **5.2** | 11.3 / **7.9** | 12.9 / **8.9** | 16.1 /**11.1** |
| RSGA | 1988-06-02 | Daily | 3 days | **Daily** | No | 4.2 | 7.8 | - | - | - |
| SEPC 3d | 1998-03-25 | Daily | 3 days | **Daily** | No | 4.1 | 7.9 | - | - | - |
| SIDC | 2000-10-16 | Daily | 3 days | **Daily** | No | 3.9 | 7.7 | - | - | - |
| KSO | 2022-12-30 | 8-Hourly | 3 days | **Daily** | **Yes** | *6.9 / 5.2 | *13.4 / 9.9 | - | - | - |
| SIFT | 2018-01-01 | **2-Hourly** | 7 days | **Daily** | **Yes** | *4.0 / 3.0 | *7.4 / 5.4 | *11.3 / 8.1 | - | - |
| SWPC 27d | 1997-06-10 | Weekly | 27 days | **Daily** | No | 4.9 | 8.5 | 12.4 | 14.1 | - |
| BGS | 2016-11-26 | Daily | 27 days | **Daily** | No | *4.7 | *8.1 | *11.8 | *12.9 | - |
| SEPC 27d | 2018-04-02 | Daily | 27 days | **Daily** | No | *4.5 | *8.5 | *13.2 | *14.3 | - |
| CLS | 1992-01-01 | Daily | 30 days | **Daily** | **Yes** | 3.9 / 3.0 | 7.2 / 5.3 | 11.5 / 8.3 | 13.3 / 9.6 | - |
| SWPC 45d | 2000-08-08 | Daily | 45 days | **Daily** | No | 4.7 | 8.1 | 12.3 | 14.3 | - |
| SWPC Solar Cycle | 2023-09-01 | Monthly | 2030-12 | Monthly | **Yes** | *26.0 / 21.1 | *25.8 / 20.6 | *26.7 /21.3 | *25.0 /20.0 | *23.3 /18.2 |
| BoM Cycle | 2000-11-02 | Monthly | Current cycle | Monthly | Not since 2007 | 15.1 / 14.2 | 15.3 / 14.4 | 16.8 /15.8 | 15.3 /14.3 | 18.7 /18.7 |
| Schatten | 1995-05-05 | Twice Yearly | ~30 years | Monthly | **Yes** | *13.1 / 9.4 | *11.1 / 7.5 | *13.6 /10.3 | *16.5 /12.4 | *20.5 /15.8 |
| MSAFE | 1999-04-01 | Monthly | Current cycle (2041) | Monthly | **Yes** | 15.6 / 12.4 | 14.9 / 11.7 | 15.2 /12.0 | 15.5 /12.3 | 18.4 /14.1 |
| Persistence | - | - | - | - | - | 4.2 | 8.6 | 14.9 | 14.5 | 25.4 |

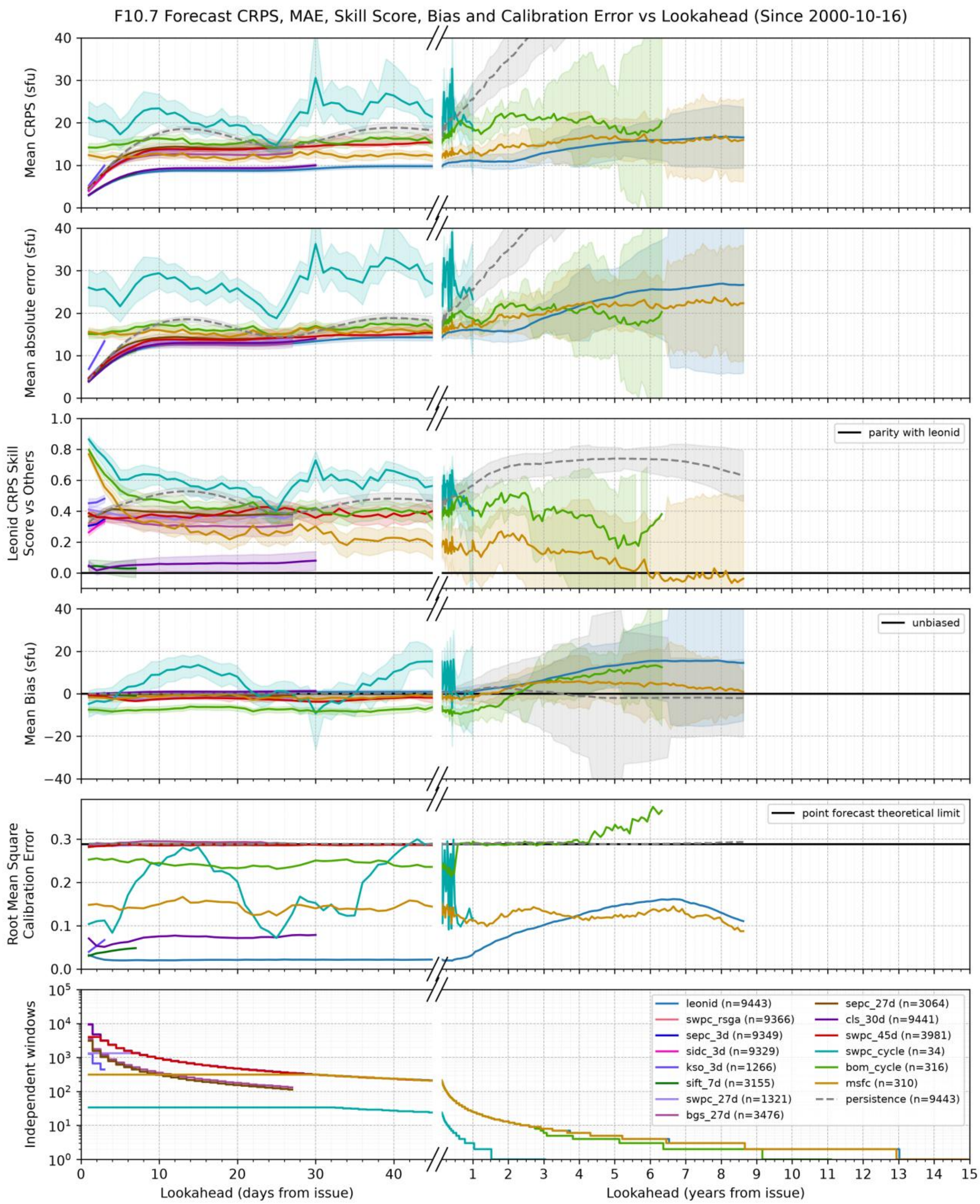

Figure 12: F10.7 forecast scores as a function of lookahead, since 2000-10-16. Note the scale change at 45 days. From top to bottom is CRPS, MAE, Leonid's SS versus the other models, mean bias, RMSCE, and the number of independent windows used for block t-intervaling the shaded confidence intervals. We assign low confidence to data with fewer than 10 windows. See the bottom subplot legend for model coloring, as well as the total number of forecast issues in this time range. Note Schatten is omitted due to a noisy signal that would obscure the plots.

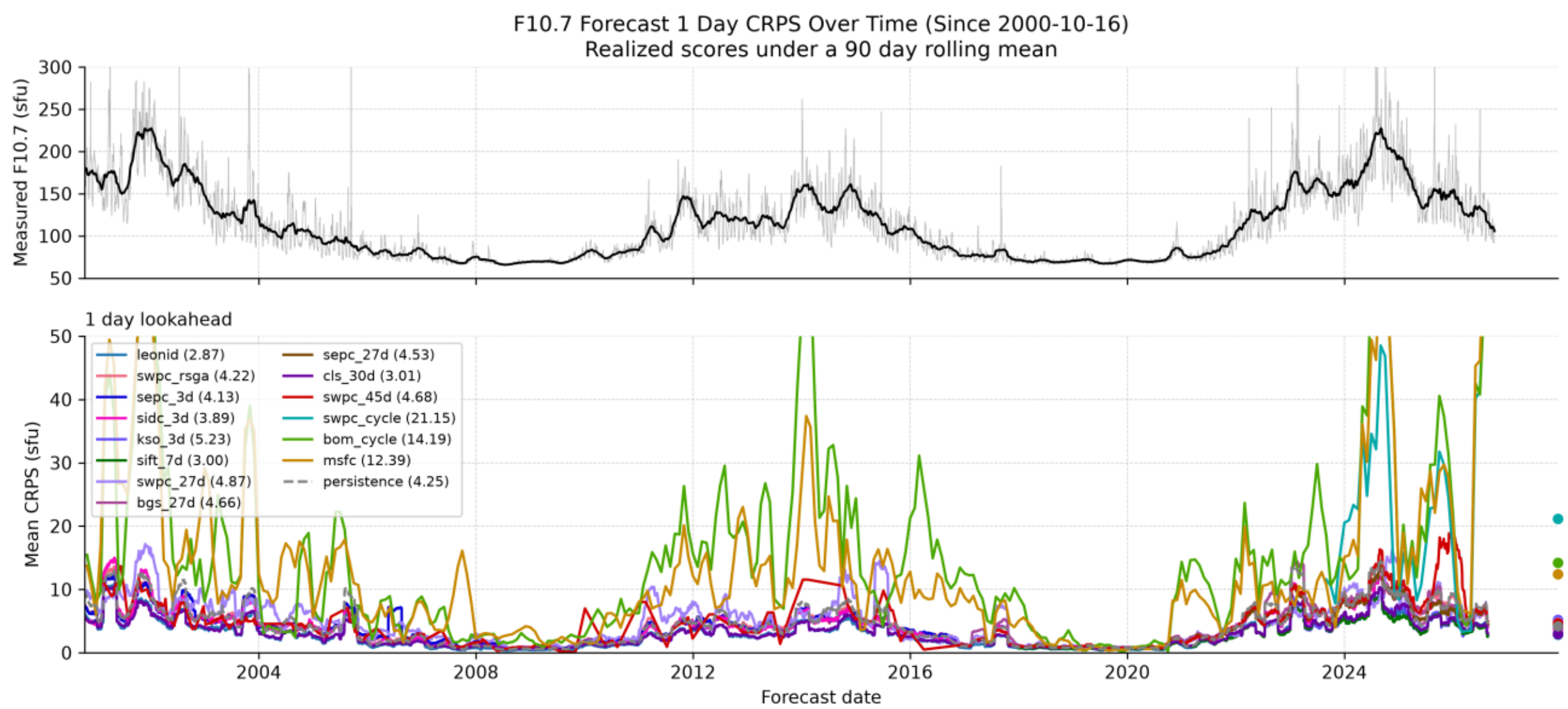

Figure 13: 90-day rolling mean of one-day lookahead CRPS for each F10.7 forecast model, showing that Leonid's performance extends across both solar minimum and maximum.

## Ap Forecast Comparisons

Table 2 shows a summary of each of the Ap forecast models, along with their MAE & CRPS at selected near-term time horizons. Leonid's forecast model scores significantly lower CRPS predictive error than all others over all time horizons, in addition to updating at the most rapid cadence to reflect changing conditions. Figure 14 shows the 1-day forecast error over time, demonstrating that the magnitude of this absolute predictive accuracy scales as a function of the solar cycle, but that Leonid's relative performance remains leading during both solar minimum and maximum. A persistence baseline is included for reference, as is a density-weighted historical average Ap of 10.

Figure 15 shows all scores for all models as a function of lookahead. Leonid's excellent CRPS performance extends smoothly over all time horizons, greatly exceeding the performance of all other forecast models at all lookaheads. This excellent CRPS scoring is due to its well-calibrated uncertainty, as demonstrated by its low RMSCE. Looking at just the MAE point forecasts, most models perform about equally as well. The mean bias plot shows Leonid's model to be unbiased, with some small bias in the remaining models. We expect this may be related to the far right-tail skew of Ap as a result of storm conditions. The drop to single-digit number of independent windows past the ~2.5-year time horizon means that behavior past this time horizon is largely characteristic of the behavior of solar cycles 23, 24, and 25, and may not generalize well.

Table 2: Comparison of Ap forecast models, along with persistence and constant historical average Ap=10 baselines. The CRPS of a point forecast is equal to its MAE, so those scores are shown as one where no uncertainty quantification exists. Bolding shows the best in each column, and asterisks indicate low-coverage scores.

| Ap Forecast Model | Archive Start | Update Cadence | Forecast Horizon | Resolution | Quantified Uncertainty | Lookahead Ap MAE / CRPS, 2000-10-16 onwards | | | | |
|---|---|---|---|---|---|---|---|---|---|---|
| | | | | | | 1-Day | 3-Day | 7-Day | 27-Day | 1-Year |
| Leonid | **1870-12-31** | **3-Hourly** | **Unbounded Multi-Cycle** | **3-Hourly** | **Yes** | 6.5 / **4.4** | 7.4 / **5.1** | 7.4 / **5.1** | 7.4 /**5.0** | 7.8 / **5.2** |
| RSGA | 1988-06-02 | Daily | 3 days | Daily | Storm Odds Only | 7.5 | 7.3 | - | - | - |
| SEPC 3d | 1998-03-25 | Daily | 3 days | Daily | No | 6.9 | 7.0 | - | - | - |
| SIDC | 2000-10-16 | Daily | 3 days | Daily | No | 6.5 | 8.1 | - | - | - |
| BoM 3d | 2000-10-19 | Daily | 3 days | Daily | No | 7.3 | 7.3 | - | - | - |
| Athens | 2014-10-01 | Daily | 3 days | Daily | No | *6.1 | *6.7 | - | - | - |
| SWPC 27d | 1997-06-10 | Weekly | 27 days | Daily | No | 7.5 | 7.1 | 7.2 | 7.1 | - |
| BGS | 2016-11-26 | Daily | 27 days | Daily | No | *6.7 | *6.7 | *7.0 | *6.7 | - |
| SEPC 27d | 2018-04-02 | Daily | 27 days | Daily | No | *6.8 | *6.8 | *6.9 | *7.0 | - |
| SWPC 45d | 2000-08-08 | Daily | 45 days | Daily | No | 7.5 | 7.2 | 6.7 | 7.4 | - |
| Schatten | 1995-05-05 | Twice Yearly | ~30 years | Monthly | No | *8.1 | *8.9 | *8.0 | *11.2 | *9.0 |
| MSAFE | 1999-04-01 | Monthly | 2041-10 | Monthly | **Yes** | 8.3 / 7.4 | 7.6 / 6.7 | 8.3 / 7.4 | 8.6 /7.7 | 8.7 / 7.5 |
| Persistence | - | - | - | - | - | 7.6 | 10.3 | 10.5 | 9.6 | 10.6 |
| Constant 10 | - | - | - | - | - | 8.1 | 8.1 | 8.1 | 8.1 | 8.1 |

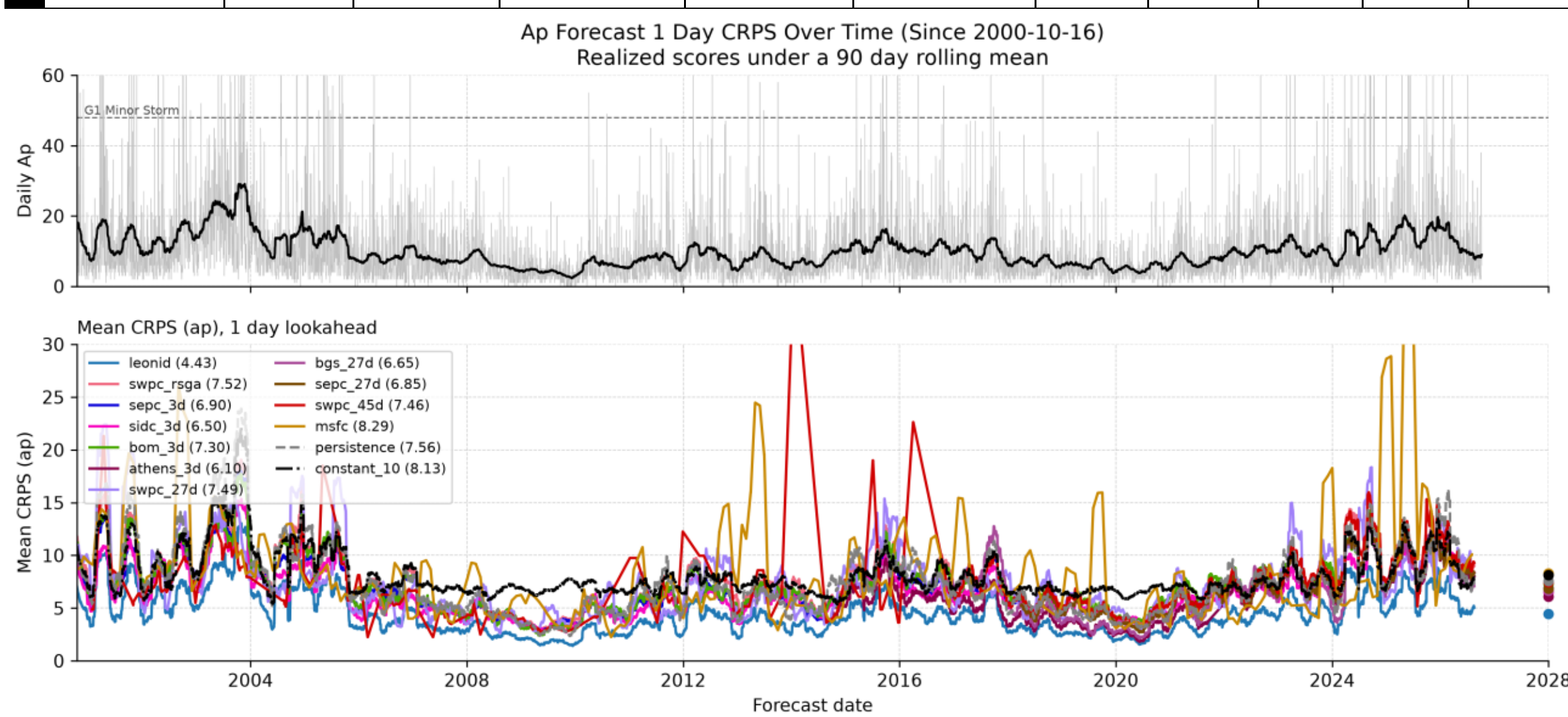

Figure 14: 90-day rolling mean of one-day lookahead CRPS for each Ap forecast model, showing that Leonid's performance extends across both solar minimum and maximum.

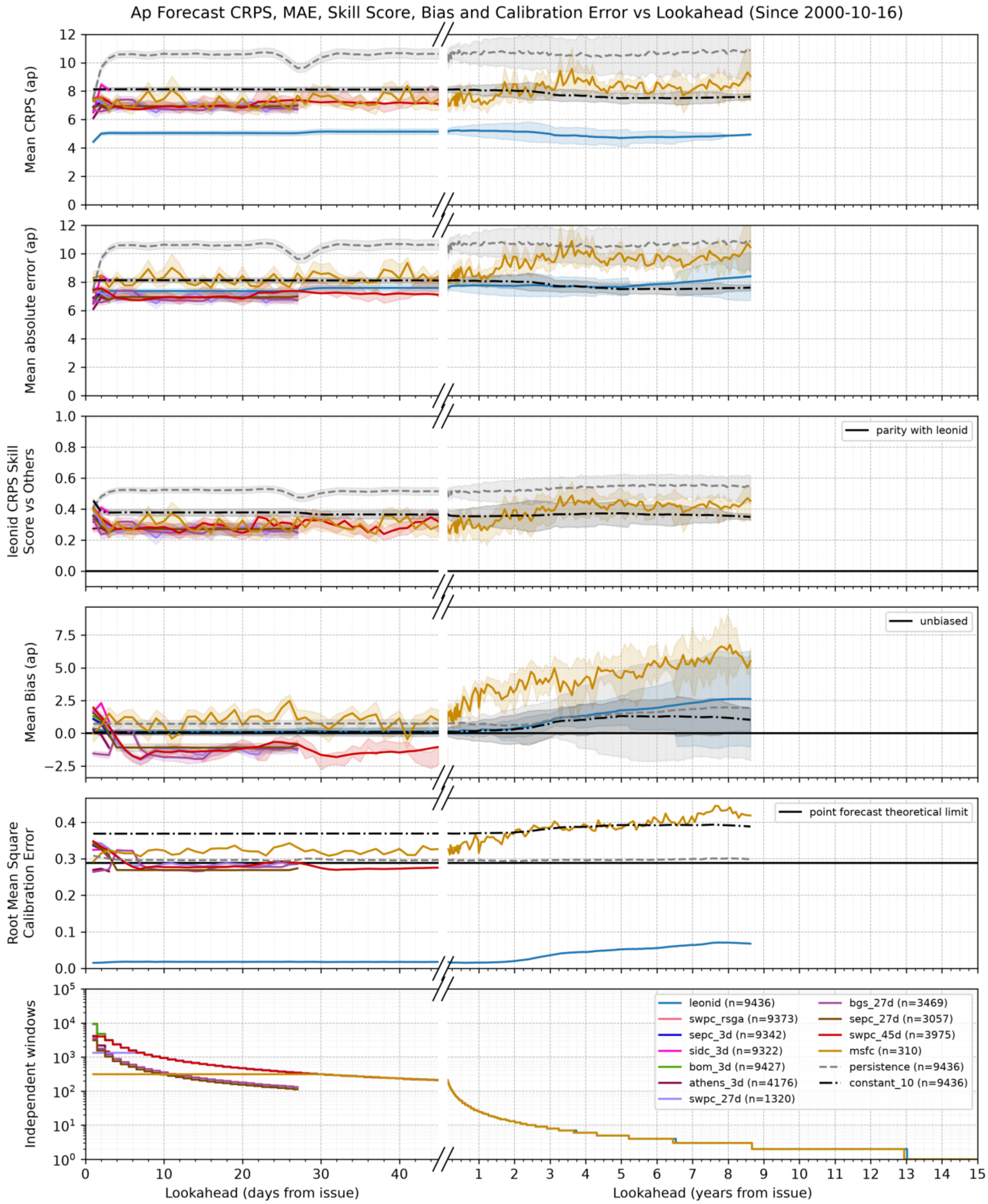

Figure 15: Ap forecast scores as a function of lookahead, since 2000-10-16. Note the scale change at 45 days. From top to bottom is CRPS, MAE, Leonid's SS versus the other models, mean bias, RMSCE, and the number of independent windows used for t-intervalling the shaded confidence intervals. We have low confidence in data with fewer than 10 windows. See the bottom subplot legend for model coloring, as well as the total number of forecast issues in this time range. Note Schatten is omitted due to a noisy signal that would obscure the plots.

## Leonid Full-Record Scoring

Figure 16 and Figure 17 below show Leonid's F10.7 and Ap LOO cross-validation scoring versus the baselines over the full space weather archives, as well as since the start of their direct measurements in 1947 and 1932, respectively. The retroforecasts prior to the start of direct F10.7 measurements have noisier base values due to the SSN conversion, which drives short-term error higher than the modern era.

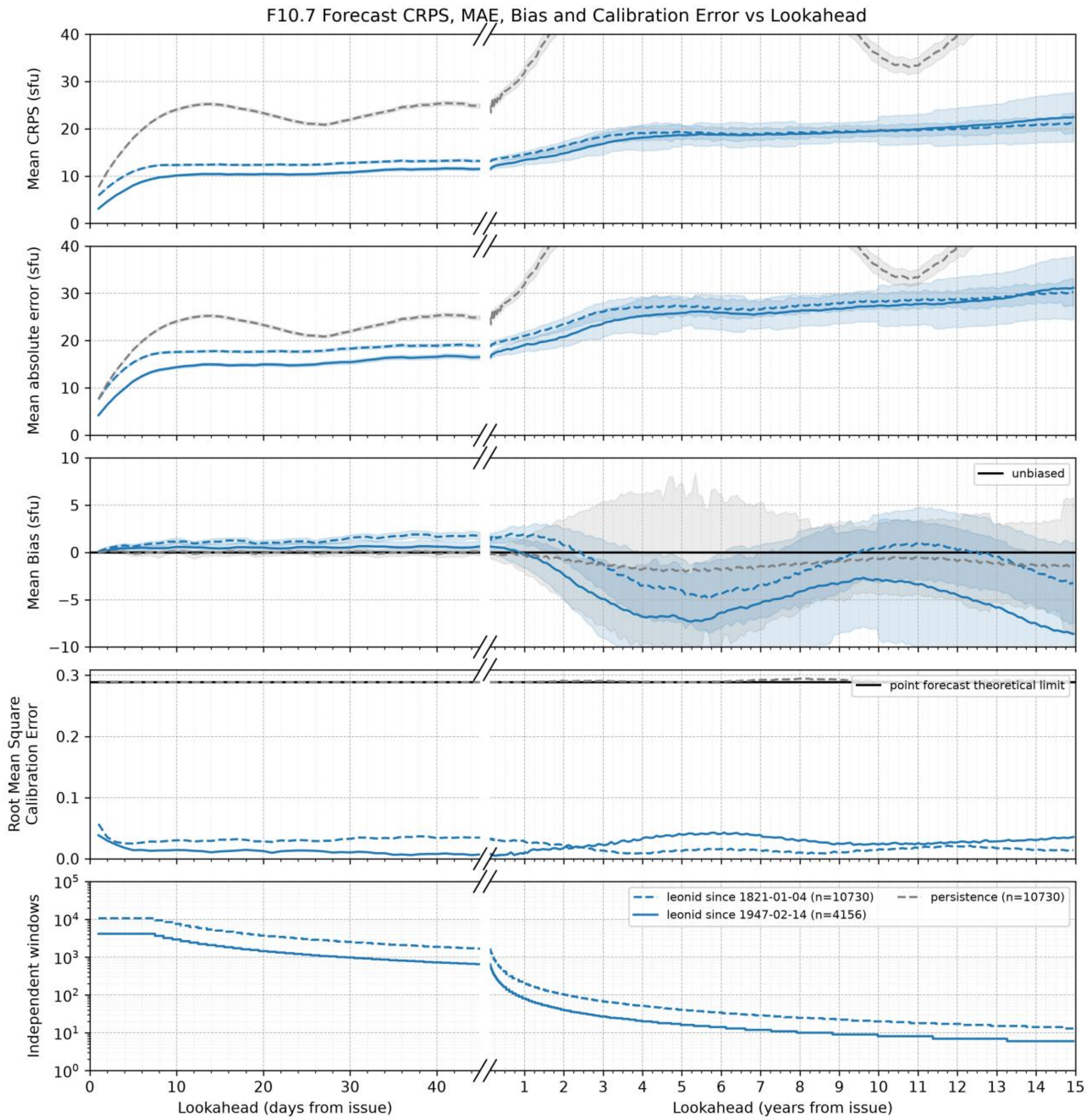

Figure 16: F10.7 forecast scores as a function of lookahead, since the start of F10.7 records on 1947-02-14 (solid), and the full SSN-converted dataset since 1821-01-04 (dashed). Note the scale change at 45 days. From top to bottom is CRPS, MAE, mean bias, RMSCE, and the number of independent windows used for t-intervalling the shaded confidence intervals.

Both F10.7 and Ap CRPS and MAE show smooth behavior over all timescales, with error slowly increasing with lookahead time. RMSCE calibration remains low, and mean bias is consistent with being unbiased over all time intervals up to 3 years. The slightly low 3-7 year bias for the since-1932 data is a result of the modern maximum, i.e. solar cycles over the past 80 years being stronger than the historical average. Future precursor measurements and predictions may be able to reduce this forecast error.

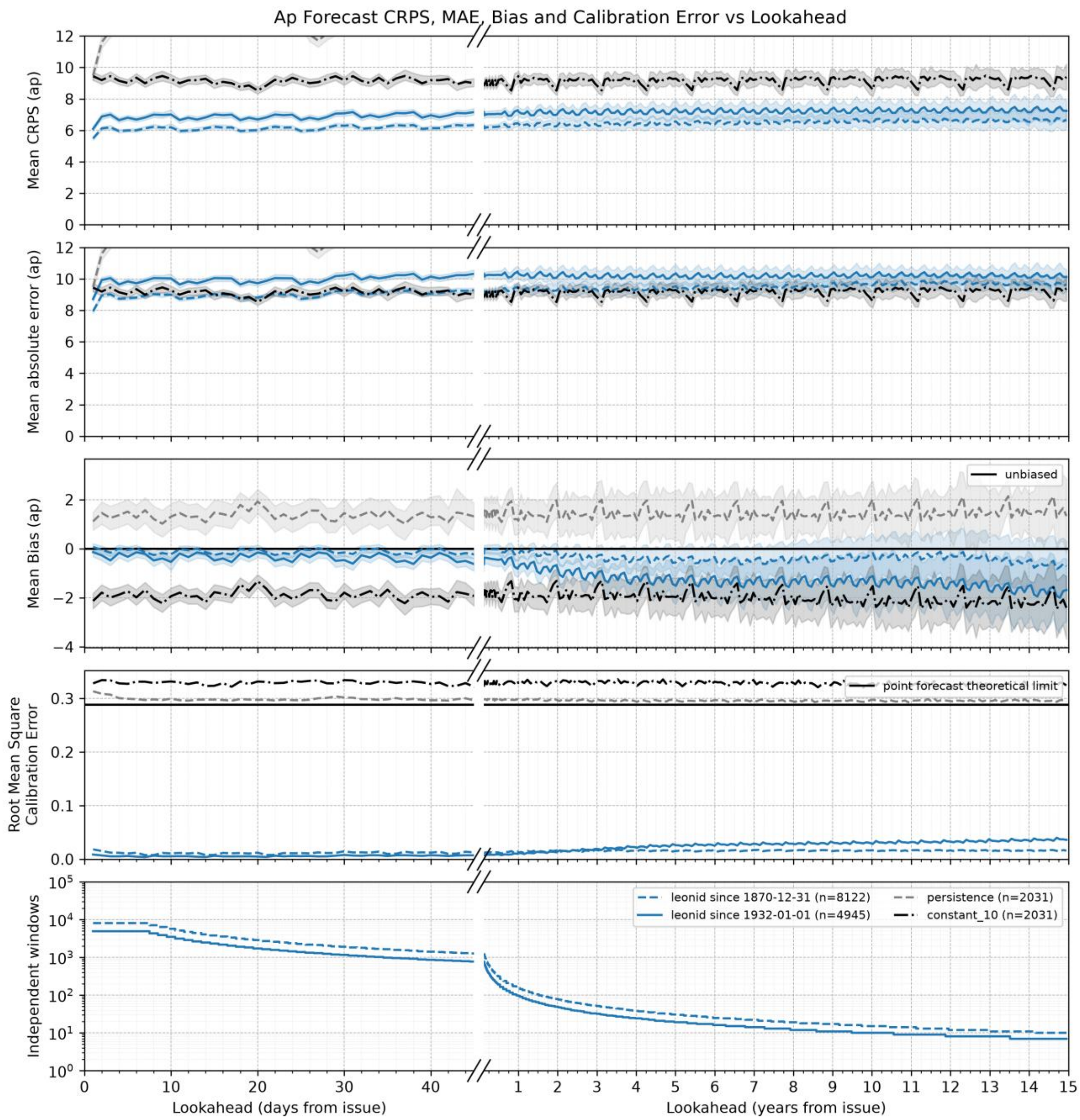

Figure 17: Ap forecast scores as a function of lookahead, since the start of Ap records in 1932-01-01 (solid), and the full aa-converted dataset since 1870-12-31 (dashed). Note the scale change at 45 days. From top to bottom is CRPS, MAE, mean bias, RMSCE, and the number of independent windows used for t-intervalling the shaded confidence intervals. Note that the periodicity in the scores is an artifact of our scoring sampling rather than the model.

## Summary

This report scored Leonid Space's F10.7 and Ap unified forecast model against all public space weather forecast archives, using a common scoring method at all lookaheads from 1 day to 15 years. Leave-one-out cross-validation was used to ensure the model was not scored on data it was fit on.

Over a common forecast interval since 2000, Leonid had the lowest CRPS prediction error at all lookaheads of any F10.7 or Ap prediction model: strictly lowest for Ap at all lookaheads and for F10.7 up to 5 years, and parity with MSAFE after 5 years. This comparison includes 15 F10.7 and 12 Ap forecasts from NASA's MSAFE, SWPC's 27-day, 45-day, and Solar Cycle predictions, SWPC/USAF's RSGA, SIDC's Ursigram, CLS, Athens, KSO, BGS, SEPC's 3-day and 27-day forecasts, AFRL/NSO's SIFT, BoM's 3-day and Cycle predictions, and NASA's Schatten forecasts. Much of this performance comes from our well-calibrated uncertainty bands and associated RMSCE scores, and we urge space weather forecasters to update their models to include calibrated uncertainties with their predictions.

These results establish a performance baseline for Leonid's space weather forecasts, and show that it has the best predictive accuracy of any forecast model compared. Our space weather indices are available via an API subscription to support space domain awareness needs such as collision avoidance, target tracking, and satellite lifetime predictions.

## References

[1] S. Shambaugh, "Validation of Satellite Lifetime Predictions at Leonid Space, Rev. 1.1," arXiv:2601.02453v2, 10 Apr 2026, https://arxiv.org/abs/2601.02453v2.

[2] J. Matzka, "The geomagnetic Kp index and derived indices of geomagnetic activity," *Space Weather,* vol. 19, no. 5, 2021.

[3] F. Clette, "Is the F 10.7 cm–Sunspot Number relation linear and stable?," *Journal of Space Weather and Space Climate,* vol. 11, 2021.

[4] K. F. Tapping and J. J. Valdes, "Did the Sun change its behaviour during the decline of cycle 23 and into cycle 24?," *Solar Physics,* 2011.

[5] NOAA Space Weather Prediction Center, "Solar Cycle 25 Forecast Update," 9 Dec 2019, https://www.spaceweather.gov/news/solar-cycle-25-forecast-update.

[6] Space Weather Prediction Center, Pushkov Institute of Terrestrial Magnetism, Ionosphere and Radio Wave Propagation (IZMIRAN), "Space Weather Forecasts," http://spaceweather.izmiran.ru/eng/forecasts.html.

[7] Institute of Ionosphere, "Geophysical Situation Forecast," https://ionos.kz/ru/uslugi/prognoz-geofizicheskoj-obstanovki/.

[8] Institute of Ionosphere, "Space Weather Forecast (archived 2016-03-22)," https://web.archive.org/web/20160322205027/http://ionos.kz/?q=en/node/21.

[9] G. Kervalishvili, I. Michaelis, M. Korte, J. Rauberg and J. Matzka, "A Novel Model for Forecasting Geomagnetic Indices Using Machine Learning," *Geophysical Research Letters,* vol. 52, no. 8, p. e2025GL114848, 2025.

[10] GFZ Helmholtz Centre for Geosciences, "Forecasting Global ap and apo Indices," https://spaceweather.gfz.de/products-data/observatory-based-geomagnetic-forecasts/forecasting-global-ap-and-apo-indices.

[11] British Geological Survey, "Predicting Space Weather," https://geomag.bgs.ac.uk/research/space_weather/bgs_predicting.html.

[12] E. Petrova, T. Podladchikova, A. M. Veronig, S. Lemmens, B. Bastida Virgili and T. Flohrer, "Medium-term Predictions of F10.7 and F30 cm Solar Radio Flux with the Adaptive Kalman Filter," *The Astrophysical Journal Supplement Series,* vol. 254, no. 1, p. 9, 2021.

[13] R. Mugellesi and D. J. Kerridge, "Prediction of Solar and Geomagnetic Activity for Low-Flying ESA Spacecraft," in *2nd International Space Debris Re-entry Workshop*, 1991.

[14] Q. Funke, A. Choudhary, D. Novak, D. Biamonti, H. Krag, S. Lemmens, T. Flohrer and V. Braun, "An Access Point to ESA's Space Debris Data: The Space Debris Office Web Based Tools," in *6th International Conference on Astrodynamics Tools and Techniques (ICATT)*, 2016.

[15] NASA Orbital Debris Program Office, "Debris Assessment Software," https://orbitaldebris.jsc.nasa.gov/mitigation/debris-assessment-software.html.

[16] W. K. Tobiska, T. Woods, F. Eparvier, R. Viereck, L. Floyd, D. Bouwer, G. Rottman and O. R. White, "The SOLAR2000 empirical solar irradiance model and forecast tool," *Journal of Atmospheric and Solar-Terrestrial Physics,* vol. 62, no. 14, pp. 1233-1250, 2000.

[17] Space Environment Technologies, "SpaceWx," https://spacewx.com/.

[18] E. Paouris, M. Abunina, A. Belov and H. Mavromichalaki, "Statistical analysis on the current capability to predict the Ap Geomagnetic Index," *New Astronomy,* vol. 86, p. 101570, 2021.

[19] NOAA Space Weather Prediction Center, "Joint USAF/NOAA Report of Solar and Geophysical Activity: README," NOAA Space Weather Prediction Center, 2008.

[20] NOAA National Centers for Environmental Information, "SWPC Products Archive: Reports of Solar Geophysical Activity," https://www.ngdc.noaa.gov/stp/space-weather/swpc-products/daily_reports/reports_solar_geophysical_activity.

[21] NOAA Space Weather Prediction Center, "Users Guide to The Preliminary Report and Forecast of Solar Geophysical Data," 2012.

[22] NOAA National Centers for Environmental Information, "SWPC Products Archive: Preliminary Report and Forecast of Solar Geophysical Data," https://www.ngdc.noaa.gov/stp/space-weather/swpc-products/weekly_reports/PRFs_of_SGD/.

[23] NOAA National Weather Service, "Service Change Notice 26-10: SWPC to Replace the USAF 45-Day Ap and F10.7 cm Flux Forecast," 2026, https://www.swpc.noaa.gov/news/update-service-change-notice-45-day-ap-and-f107-forecast.

[24] Internet Archive, "Wayback Machine," https://web.archive.org/.

[25] ESA Space Weather Service Network, Institute of Space Science (Romania), "FORIND: Nowcast and Forecast Indices for Atmospheric Drag Calculation," https://swe.ssa.esa.int/web/guest/forind-federated.

[26] NOAA National Centers for Environmental Information, "SWPC Products Archive: 45-Day Forecast," https://www.ngdc.noaa.gov/stp/space-weather/swpc-products/daily_reports/45_day_forecast.

[27] D. Berghmans, R. A. M. Van der Linden, P. Vanlommel, R. Warnant, A. Zhukov, E. Robbrecht, F. Clette, O. Podladchikova, B. Nicula, J.-F. Hochedez, L. Wauters and S. Willems, "Solar Activity: Nowcasting and Forecasting at the SIDC," *Annales Geophysicae,* vol. 23, pp. 3115-3128, 2005.

[28] Solar Influences Data Analysis Center, Royal Observatory of Belgium, "Space Weather Services Archive: URSIgram (meu)," https://www.sidc.be/spaceweatherservices/managed/services/archive/product/meu.

[29] Collecte Localisation Satellites, "Solar Radio Flux," https://spaceweather.cls.fr/services/radioflux/ (forecast files at ftp://ftpsedr.cls.fr/pub/previsol/solarflux/forecast/absolute).

[30] P. Yaya, L. Hecker, T. Dudok de Wit, C. Le Fèvre and S. Bruinsma, "Solar Radio Proxies for Improved Satellite Orbit Prediction," *Journal of Space Weather and Space Climate,* vol. 7, p. A35, 2017.

[31] National Solar Observatory and Air Force Research Laboratory, "ADAPT-GONG magSums SIFT Dashboard," https://gong.nso.edu/ftp/adapt/sift/adapt-sift_dashboard-readme_nso.html.

[32] C. J. Henney, W. A. Toussaint, S. M. White and C. N. Arge, "Forecasting F10.7 with solar magnetic flux transport modeling," *Space Weather,* vol. 10, no. 2, p. S02011, 2012.

[33] C. N. Arge, C. J. Henney, J. Koller, C. R. Compeau, S. Young, D. MacKenzie, A. Fay and J. W. Harvey, "Air Force Data Assimilative Photospheric Flux Transport (ADAPT) Model," *AIP Conference Proceedings,* vol. 1216, no. 1, pp. 343-346, 2010.

[34] Athens Space Weather Forecasting Center, National and Kapodistrian University of Athens, "Daily Forecast of Geomagnetic Activity," http://spaceweather.phys.uoa.gr/.

[35] ESA Space Weather Service Network, "KSO Solar Radio Flux Forecast (S107h)," https://swe.ssa.esa.int/web/guest/kso-S107h-federated.

[36] O. Podladchikova, C. Marque, T. Podladchikova, A. Veronig and K. Stegen, "Comparison of Different Approaches for the Real Time Forecast of the F10.7 index at the Space Weather Forecasting Centres," in *EGU General Assembly*, 2019.

[37] K. F. Tapping, "The 10.7 cm solar radio flux (F10.7)," *Space Weather,* vol. 11, no. 7, pp. 394-406, 2013.

[38] Space Environment Prediction Center, National Space Science Center, Chinese Academy of Sciences, "Archive: Daily Reports," http://www.sepc.ac.cn/eng/archive.php.

[39] Space Environment Prediction Center, National Space Science Center, Chinese Academy of Sciences, "27-day F10.7 and Ap Forecasts," http://www.sepc.ac.cn/eng/F107Forecast.php and http://www.sepc.ac.cn/eng/ApForecast.php.

[40] S. Liu, B. Luo, Q. Zhong and X. Dou, "Medium and short term forecasting of Ap related to coronal holes," *Chinese Journal of Space Science,* vol. 29, no. 6, pp. 545-551, 2009.

[41] British Geological Survey, "Space Weather Forecast," https://geomag.bgs.ac.uk/data_service/space_weather/forecast.html.

[42] Australian Bureau of Meteorology, Australian Space Weather Forecasting Centre, "ips-dsgr: Daily Solar and Geophysical Report," https://listserver.ips.gov.au/pipermail/ips-dsgr/.

[43] Australian Bureau of Meteorology, Australian Space Weather Forecasting Centre, "ips-ssn-predictions: Observed and Predicted Solar Indices," https://listserver.ips.gov.au/pipermail/ips-ssn-predictions/.

[44] M. Miesch, "Solar Cycle Progression Updated Prediction (Experimental): Validation Document," NOAA Space Weather Prediction Center, 2024.

[45] NOAA National Centers for Environmental Information, "SWPC Products Archive: Solar Cycle Testbed Predictions," https://www.ngdc.noaa.gov/stp/space-weather/swpc-products/SCtestbed/.

[46] NOAA National Centers for Environmental Information, "SWPC Products Archive: Solar Cycle Progression," https://www.ngdc.noaa.gov/stp/space-weather/swpc-products/solar_cycle_progression.

[47] M. S. Miesch, "Solar cycle prediction at NOAA's space weather prediction center," *Space Weather,* vol. 23, no. 6, p. e2025SW004444, 2025.

[48] NOAA Space Weather Prediction Center, "Solar Cycle Progression," https://www.swpc.noaa.gov/products/solar-cycle-progression.

[49] NASA Marshall Space Flight Center, "Solar Cycle Progression and Forecast: Current and Archived Forecasts," https://www.nasa.gov/solar-cycle-progression-and-forecast/ and https://www.nasa.gov/solar-cycle-progression-and-forecast/archived-forecast/.

[50] K. O. Niehuss, H. C. Euler Jr. and W. W. Vaughan, "Statistical Technique for Intermediate and Long-Range Estimation of 13-Month Smoothed Solar Flux and Geomagnetic Index," NASA, 1996.

[51] R. J. Suggs, "Future Solar Activity Estimates for Use in Prediction of Space Environmental Effects on Spacecraft Orbital Lifetime and Performance," National Aeronautics and Space Administration, George C. Marshall Space Flight Center, Huntsville, 2016.

[52] A. G. McNish and J. V. Lincoln, "Prediction of Sunspot Numbers," *Eos, Transactions American Geophysical Union,* vol. 30, no. 5, pp. 673-685, 1949.

[53] F. Clette, S. Jain and T. Podladchikova, "The McNish and Lincoln Solar-Activity Predictions: The Method and Its Performance," *Solar Physics,* vol. 299, no. 22, 2024.

[54] NASA GSFC Flight Dynamics Facility, "SchattenSolarFluxPrediction," https://iswa.gsfc.nasa.gov/iswa_data_tree/model/solar/FDF/SchattenSolarFluxPrediction/.

[55] K. H. Schatten, P. H. Scherrer, L. Svalgaard and J. M. Wilcox, "Using dynamo theory to predict the sunspot number during solar cycle 21," *Geophysical Research Letters,* vol. 5, no. 5, pp. 411-414, 1978.

[56] K. Schatten, "Fair space weather for solar cycle 24," *Geophysical Research Letters,* vol. 32, no. 21, p. L21106, 2005.

[57] R. Ibragimov and U. K. Müller, "t-Statistic Based Correlation and Heterogeneity Robust Inference," *Journal of Business & Economic Statistics,* vol. 28, no. 4, pp. 453-468, 2010.

[58] E. C. Fieller, "Some Problems in Interval Estimation," *Journal of the Royal Statistical Society Series B: Statistical Methodology,* vol. 16, no. 2, pp. 175-185, 1954.